\documentclass[useAMS,usenatbib,usegraphicx]{mn2e}

\usepackage{color}
\usepackage{natbib, aas_macros}
\title[Central Mass Profiles of Hydra A and A478]{Central Mass Profiles
of the Nearby Cool-core Galaxy Clusters Hydra A and A478\thanks{Based on data collected at Subaru Telescope, which is operated by the National Astronomical Observatory of Japan.}}
\author[N. Okabe et al.]
{N. Okabe $^{1,2,3,4}$\thanks{E-mail:okabe@hiroshima-u.ac.jp},
K. Umetsu$^{4}$,
T. Tamura$^{5}$,
Y. Fujita$^{6}$,
M. Takizawa$^{7}$, 
\newauthor
K. Matsushita$^{8}$,
Y. Fukazawa$^{2}$, 
T. Futamase$^{9}$,
M. Kawaharada$^{5}$,
\newauthor
S. Miyazaki$^{10}$,
Y. Mochizuki$^{8}$,
K. Nakazawa$^{11}$,
T. Ohashi$^{12}$,
N. Ota$^{13}$,
\newauthor
T. Sasaki$^{8}$,
K. Sato$^{8}$,
and
S. I. Tam$^{14}$\\
$^{1}$ Department of Physical Science, Hiroshima University, 1-3-1
Kagamiyama, Higashi-Hiroshima, Hiroshima 739-8526, Japan\\
$^{2}$Hiroshima Astrophysical Science Center, Hiroshima University, Higashi-Hiroshima,
Kagamiyama 1-3-1, 739-8526, Japan\\
$^{3}$ Kavli Institute for the Physics and Mathematics of the Universe (WPI), \\Todai Institutes for Advanced Study,University of Tokyo, 5-1-5 Kashiwanoha, Kashiwa, Chiba 277-8583, Japan \\
$^{4}$Institute of Astronomy and Astrophysics, Academia Sinica, P.O. Box 23-141, Taipei 10617, Taiwan\\
$^{5}$Institute of Space and Astronautical Science, Japan Aerospace Exploration Agency,\\ 3-1-1 Yoshinodai, Chuo-ku, Sagamihara, Kanagawa 229-8510, Japan \\
$^{6}$Department of Earth and Space Science, Graduate School of Science, Osaka University, Toyonaka, Osaka 5
60-0043, Japan\\
$^{7}$Department of Physics, Yamagata University, Kojirakawa-machi 1-4-12, Yamagata 990-8560, Japan \\
$^{8}$Department of Physics, Tokyo University of Science, 1-3 Kagurazaka, Shinjyuku-ku, Tokyo 162-8601, Japan\\
$^{9}$Astronomical Institute, Tohoku University, Aramaki, Aoba-ku, Sendai 980-8578, Japan\\
$^{10}$National Astronomical Observatory of Japan, Mitaka, Tokyo 181-8588, Japan\\
$^{11}$Department of Physics, The University of Tokyo, 7-3-1 Hongo, Bunkyo-ku, Tokyo 113-0033, Japan\\
$^{12}$Department of Physics, Tokyo Metropolitan University, 1-1 Minami-Osawa, Hachioji, Tokyo 192-0397, Japan\\
$^{13}$Department of Physics, Nara Women's University, Kitauoyanishi-machi, Nara, Nara 630-8506, Japan \\
$^{14}$Department of Physics ,National Taiwan University, Taipei 10617, Taiwan \\
}

\begin{document}

\date{\today}

\pagerange{\pageref{firstpage}--\pageref{lastpage}} \pubyear{2002}

\maketitle

\label{firstpage}

\newcommand{\simgt}{\lower.5ex\hbox{$\; \buildrel > \over \sim \;$}}
\newcommand{\simlt}{\lower.5ex\hbox{$\; \buildrel < \over \sim \;$}}
\newcommand{\tp}{\hspace{-1mm}+\hspace{-1mm}}
\newcommand{\tm}{\hspace{-1mm}-\hspace{-1mm}}
\newcommand{\gIII}{I\hspace{-.3mm}I\hspace{-.3mm}I}
\newcommand{\bmf}[1]{\mbox{\boldmath$#1$}}
\def\bbeta{\mbox{\boldmath $\beta$}}
\def\btheta{\mbox{\boldmath $\theta$}}
\def\bnabla{\mbox{\boldmath $\nabla$}}
\def\bk{\mbox{\boldmath $k$}}
\newcommand{\cA}{{\cal A}}
\newcommand{\cD}{{\cal D}}
\newcommand{\cF}{{\cal F}}
\newcommand{\cG}{{\cal G}}
\newcommand{\trQ}{{\rm tr}Q}
\newcommand{\Real}[1]{{\rm Re}\left[ #1 \right]}
\newcommand{\paren}[1]{\left( #1 \right)}
\newcommand{\red}{\textcolor{red}}
\newcommand{\blue}{\textcolor{blue}}
\newcommand{\norv}[1]{{\textcolor{blue}{#1}}}
\newcommand{\norvnew}[1]{{\textcolor{red}{#1}}}

\def\hkpc{\mathrel{h^{-1}{\rm kpc}}}
\def\hMpc{\mathrel{h^{-1}{\rm Mpc}}}
\def\Mvir{\mathrel{M_{\rm vir}}}
\def\cvir{\mathrel{c_{\rm vir}}}
\def\rvir{\mathrel{r_{\rm vir}}}
\def\Dvir{\mathrel{\Delta_{\rm vir}}}
\def\rsc{\mathrel{r_{\rm sc}}}
\def\rhoc{\mathrel{\rho_{\rm crit}}}
\def\Msol{\mathrel{M_\odot}}
\def\hMsol{\mathrel{h^{-1}M_\odot}}
\def\h70Msol{\mathrel{h_{70}^{-1}M_\odot}}

\begin{abstract}
We perform a weak-lensing study of the nearby cool-core galaxy clusters, 
Hydra A ($z=0.0538$) and A478 ($z=0.0881$), 
of which brightest cluster galaxies (BCGs) host powerful activities of
 active galactic nuclei (AGNs).
For each cluster, the observed tangential shear profile is well
 described either by a single Navarro--Frenk--White model or a two-component
 model including the BCG as an unresolved point mass.
For A478, we determine the BCG and its host-halo masses from a joint fit to
 weak-lensing and stellar photometry measurements.
We find that the choice of initial mass functions (IMFs) 
 can introduce a factor of two uncertainty in the BCG mass, whereas the
 BCG host halo mass is well constrained by data. 
We perform a joint analysis of weak-lensing and stellar kinematics data available
 for the Hydra A cluster, which allows us to constrain the central mass
 profile without assuming specific IMFs.
We find that the central mass profile ($r<300$\,kpc) determined from
the joint analysis is
 in excellent agreement with those from independent measurements, 
 including
 dynamical masses estimated from the cold gas disk component,
 X-ray hydrostatic total mass estimates, 
 and 
 the central stellar mass estimated based on the Salpeter IMF. 
The observed dark-matter fraction around the BCG for Hydra A is found
to be smaller than those predicted by adiabatic contraction
models, suggesting the importance of other physical processes, such as
the the AGN feedback and/or dissipationless
mergers.
\end{abstract}


\makeatletter

\def\doi{\begingroup
  % The following isn't just \dospecials, because that includes \ , \{, and \}
  \let\do\@makeother \do\\\do\$\do\&\do\#\do\^\do\_\do\%\do\~
  \@ifnextchar[%]
    {\@doi}
    {\@doi[]}}
\def\@doi[#1]#2{%
  \def\@tempa{#1}%
  \ifx\@tempa\@empty
    \href{http://dx.doi.org/#2}{doi:#2}%
  \else
    \href{http://dx.doi.org/#2}{#1}%
  \fi
  \endgroup
}

%
\def\eprint#1#2{%
  \@eprint#1:#2::\@nil}
\def\@eprint@arXiv#1{\href{http://arxiv.org/abs/#1}{{\tt arXiv:#1}}}
\def\@eprint@dblp#1{\href{http://dblp.uni-trier.de/rec/bibtex/#1.xml}{dblp:#1}}
\def\@eprint#1:#2:#3:#4\@nil{%
  \def\@tempa{#1}%
  \def\@tempb{#2}%
  \def\@tempc{#3}%
  \ifx\@tempc\@empty
    \let\@tempc\@tempb
    \let\@tempb\@tempa
  \fi
  \ifx\@tempb\@empty
    \def\@tempb{arXiv}%
  \fi
  \@ifundefined{@eprint@\@tempb}
    {\@tempb:\@tempc}
    {
      \expandafter\expandafter\csname @eprint@\@tempb\endcsname\expandafter{\@tempc}}%
}

%
\def\mniiiauthor#1#2#3{%
  \@ifundefined{mniiiauth@#1}
    {\global\expandafter\let\csname mniiiauth@#1\endcsname\null #2}
    {#3}}

\makeatother

\section{Introduction}

The standard cold-dark-matter (CDM) paradigm excellently
describes the statistical properties and growth of cosmic structure 
from the early universe to the present day epoch,
as observed from the cosmic microwave background radiation
\citep[e.g.][]{WMAP09,Planck13Cosmology} and large scale clustering of
galaxies
 \citep[e.g.][]{2DF,Anderson14}.
Cosmological simulations of collisionless CDM predict that
the quasi-equilibrium mass density profile of dark-matter halos
formed by collisionless gravitational dynamics
is nearly universal and self-similar
over wide mass ranges \citep[][hereafter NFW]{NFW96}.
The spherically-averaged density profile exhibits a soft cusp ($\rho \propto r^{-1}$) 
in the center
and has steeper slopes at larger halo radii
 ($\rho \propto r^{-3}$ in the outskirts).
Recent observations of stacked weak gravitational lensing signals
for massive galaxy clusters at $z\sim0.2$ 
\citep{Oguri12,Okabe10b,Okabe13,Okabe15c,Umetsu11,Umetsu14,Umetsu15b,Niikura15} 
revealed 
that the observed cluster 
lensing profiles are in excellent agreements with the NFW density
profile predicted for collisionless CDM halos.
However, since the innermost cluster radius for weak lensing is
typically limited to $r\sim100h^{-1}$\,kpc 
due to the finite resolution of weak-lensing observations,
it is practically difficult to directly constrain the innermost cluster
mass distribution around bright cluster galaxies (BCGs).

The total mass density profiles of clusters in the central region
have been a subject of intense theoretical and observational studies.
Numerical simulations of collisionless dark matter show
that the central cusp slope of dark-matter halos is not strictly self
similar \citep{Gao12}.
Baryonic effects cannot be ignored in the central region where the
cooling time is shorter than the Hubble time and rich stellar
populations within BCGs are abundant.
Stars form from radiative gas cooling,
and subsequently dark matter is pulled inward in response to the central
condensation of baryons.
A cuspy dark-matter density profile is formed through 
such energy exchange processes
between dark matter and baryons, 
which is refereed to as {\em adiabatic contraction}
\citep[e.g.][]{Eggen62,Blumenthal86,Gao04c,Gnedin04}.
On the other hand, if BCGs are formed by the accretion of 
clumpy structures such as satellite galaxies and globular clusters,
a cored density profile is expected to form by dynamical friction
\citep[e.g.][]{El-Zant01,Lackner10}. 
Gas outflows driven by active galactic nuclei (AGNs) 
can suppress condensation of gas and stars,
resulting in flattening of the dark-matter distribution in the central
region \citep[e.g.][]{Teyssier11,Martizzi12,Ragone-Figueroa12}.

Stellar kinematics of BCGs is one of the most powerful means of
measuring the total mass profile in the central region of galaxies
\citep[e.g.][]{Cappellari06}. 
Recently, \citet{Newman09,Newman11,Newman13} measured 
the total mass profiles of relaxed clusters and decomposed them into
 into stellar and dark-matter components
from joint measurements of strong/weak lensing and stellar
kinematics within BCGs,
under the assumptions of dynamical equilibrium, isotropic orbits, and
constant stellar mass-to-light ratio.
\cite{Oguri14} carried out a joint ensemble analysis of strong lensing and
stellar photometry data for a large sample of 161 elliptical galaxies
to measure their average mass density profile.
Stellar mass estimates for galaxies can be obtained from stellar
photometry using scaling relations between galaxy luminosity and stellar
mass, whereas the stellar initial mass function (IMF) governs the
normalization of the stellar mass to luminosity relation.
Hence, the uncertainty of the stellar IMF can give
rise to a systematic bias in stellar mass estimates.
In order to break the degeneracy between the IMF uncertainty and 
the inner slope of the mass density profile, \cite{Oguri14} proposed to use quasar
microlensing observations. 


We have two ideal clusters,
Hydra A \citep[$z=0.0538$;][]{Sato12} and Abell 478
\citep[$z=0.0881$;][]{Mochizuki14}, to study the mass distributions
within both clusters and BCGs.
The two clusters have been selected and observed as targets of joint
Subaru weak-lensing and {\em Suzaku} X-ray studies
\citep{Okabe14b}.
They are well-known cool-core clusters 
hosting central AGNs inside their BCGs.
Recent X-ray and radio observations
\citep[e.g.][]{McNamara00,David01,Wise07,Sun03,Sanderson05,Diehl08}
have shown that giant radio lobes triggered by prominent AGN jets are 
interacting with the surrounding intracluster medium (ICM). 
Isobaric cooling time scales in the central region are shorter than
$1$\,Gyr.
Hence, the clusters provide us with unique environments for studying
the effects of baryons acting on the dark-matter distribution in the
cluster center.

Since these clusters are at very low redshifts, 
their large angular extents enable us to detect weak-lensing signals
around the BCGs, in particular for Hydra A, and to measure the radial
mass distribution, by using the unique combination of weak lensing and stellar
kinematics or stellar photometry.
For these very nearby clusters, especially for the Hydra A cluster,
strong-lensing information is not needed to resolve the central matter distribution.

This paper is organized as follows.
In Section \ref{sec:data}, we conduct shape measurements, 
examine the redshift dependence of the contamination level by member
galaxies, and securely select background galaxies.
We measure cluster masses in Section \ref{sec:WLmass}, perform joint
analyses in Section \ref{sec:dis}, and discuss the results
 in Section \ref{sec:dis2}.
Section \ref{sec:sum} is devoted to the summary.
A comparison of weak-lensing masses and {\it Suzaku} X-ray observables 
is given in our previous paper \citep{Okabe14b}.
We use $\Omega_{m,0}=0.27$, $\Omega_{\Lambda}=0.73$ and 
$H_0=70h_{70}$\,km\,s$^{-1}$\,Mpc$^{-1}$.

\section{Data Analysis} \label{sec:data}

\subsection{Data Analysis and Shape Measurement}

\begin{figure*}
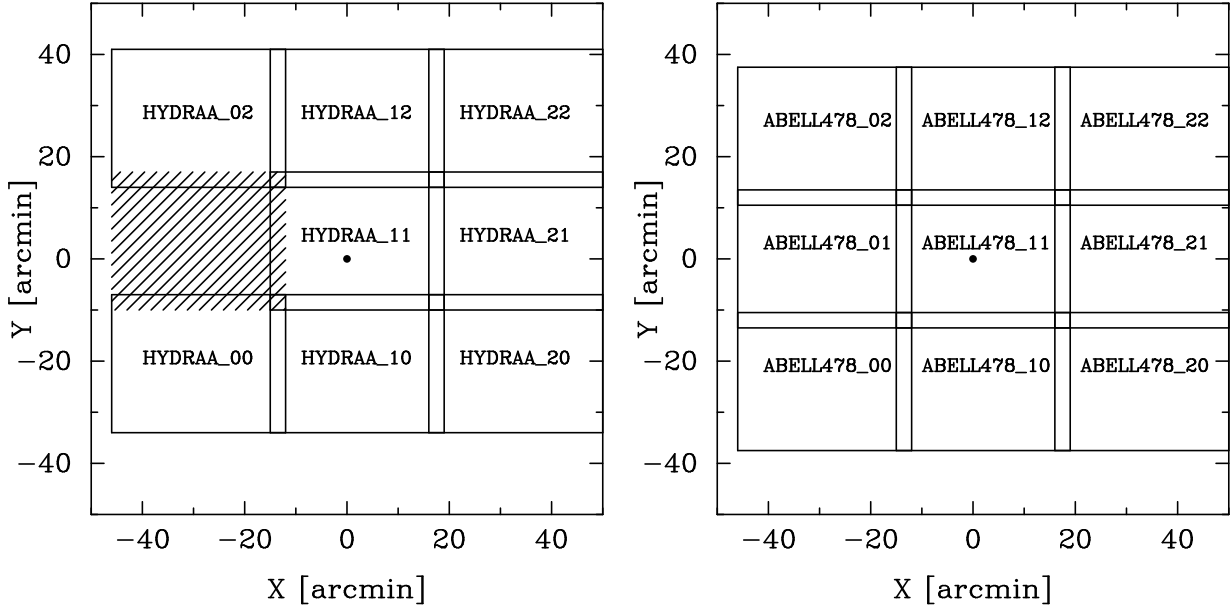

\begin{center}$
\begin{array}{cc}
\includegraphics[width=0.45\hsize]{f1a.ps} &
\includegraphics[width=0.45\hsize]{f1b.ps}
\end{array}$
\end{center}
\caption{Pointing maps for Hydra A (left) and Abell 478 (right).
The horizontal and vertical axes are R.A. and Decl. offsets from BCGs, 
in units of arcmin.
For each cluster, the circle represents the position of the BCG:
($139.524^\circ$,$-12.095^\circ$) for Hydra A and 
($63.355^\circ$,$10.465^\circ$) for A478.
The pointing IDs (Table \ref{tab:data}) are shown for each cluster.
The adjacent pointings have an overlap of 2 arcmin. 
The hatched region indicates that there is no data for the present study.
}
\label{fig:pointing}
\end{figure*}

We conducted observations of 
Hydra A and Abell 478 
 using the Subaru/Suprime-cam \citep{Miyazaki02}, in $i'$ and $g'$
 bands, on Jan 7th, 2013.
The $i$ imaging was taken to measure ellipticities of galaxies for the
weak-lensing shape analysis.
The two-band photometry in the $g'$ and $i'$ bands is
to minimize contamination of the background source catalog 
by unlensed cluster member galaxies.
The Hydra A and Abell 478 fields are
covered by a mosaic of 8 and 9 pointings,
respectively, as shown in Figure \ref{fig:pointing}. 
We note that, for Hydra A, the middle-east pointing was not available as
limited by the number of allocated nights and bright stars in that region.
The observing conditions are summarized in Table \ref{tab:data}. 
The typical seeing is $\sim 0.7\arcsec$ in the $i'$ band for
weak-lensing shape measurements.
The exposure times for the $i'$ and $g'$ bands are $\sim15$ and $\sim12$\,min, respectively.
These are about half of those typically used in our previous weak-lensing studies
of intermediate-redshift clusters at $z\sim 0.2$ 
\citep[e.g.,][]{Okabe08,Okabe10b,Okabe13,Okabe15c}.
We used the standard Suprime-Cam reduction software
{\sc SDFRED} \citep{Yagi02,Ouchi04} modified to accommodate new CCD chips, 
for flat-fielding, instrumental distortion correction, differential
refraction, point-spread-function (PSF) matching, sky subtraction and stacking. 
An astrometric calibration was conducted with 2MASS point sources
 \citep{2MASS06}.
Typical residual astrometric offsets are no larger than the CCD pixel size.

\begin{table}
\caption{Summary of Subaru observations for Hydra A and Abell 478. 
Each pointing is shown in Figure \ref{fig:pointing}.
The seeing FWHM is for the $i'$-band used for shape measurements.
} \label{tab:data}
\begin{center}
\begin{tabular}{cccc}
\hline
\hline
Name & $i'$
     & $g'$
     & Seeing \\
     & [min]
     & [min]
     & [arcsec] \\
\hline
HYDRAA\_00 & 15.0
         & 11.7
         & 0.75\\
HYDRAA\_02 & 15.0
         & 10.0
         & 0.71\\
HYDRAA\_10 & 15.0
         & 11.7
         & 0.73\\
HYDRAA\_11 & 15.0
         & 11.2
         & 0.71\\
HYDRAA\_12 & 15.0
         & 10.0
         & 0.63\\
HYDRAA\_20 & 15.0
         & 11.7
         & 0.67\\
HYDRAA\_21 & 15.0
         & 11.2
         & 0.71\\
HYDRAA\_22 & 15.0
         & 10.0
         & 0.65\\
\hline
ABELL478\_00 & 15.0
         & 11.7
         & 0.67\\
ABELL478\_01 & 15.0
         & 11.7
         & 0.79\\
ABELL478\_02 & 15.0
         & 11.7
         & 0.73\\
ABELL478\_10 & 15.0
         & 11.7
         & 0.65\\
ABELL478\_11 & 15.0
         & 11.7
         & 0.79\\
ABELL478\_12 & 15.0
         & 11.7
         & 0.71\\
ABELL478\_20 & 15.0
         & 14.0
         & 0.69\\
ABELL478\_21 & 15.0
         & 11.7
         & 0.69\\
ABELL478\_22 & 15.0
         & 11.7
         & 0.71\\
\hline
\end{tabular}
\end{center}
\end{table}

We determine the mass distribution in nearby clusters using weak
gravitational lensing.
Weak lensing analysis is performed following \cite{KSB} with some
modifications \citep{Okabe13,Okabe14a,Okabe15c}. 
Technical details are described in \cite{Okabe14a}.
The image ellipticity $e_\alpha$ of an object detected in the $i'$ band data
is measured from weighted quadrupole moments of the surface brightness
distribution.
The PSF anisotropy is corrected for using second-order bipolynomial
functions of the stellar anisotropy kernel, which is expressed with
KSB's smear polarizability tensor and image ellipticity of unsaturated
stars. 
To examine the validity of anisotropic PSF corrections,
we have checked residual systematics in corrected ellipticities as
described in Appendix of \cite{Okabe14a}. 
Two-point correlation functions between star and galaxy ellipticities
after the correction are of the order of $10^{-8}$--$10^{-6}$, 
significantly improved from $10^{-5}$--$10^{-4}$ before the correction.
This level of residual correlations is consistent with zero.
The reduced shear signal is obtained by applying a correction
for the isotropic smearing effect as
$g_\alpha=(P_g^{-1})_{\alpha\beta} e_{\beta}'$.
Here, $e_\alpha'$ is the image ellipticity after the anisotropic PSF correction
and $P^g_{\alpha\beta}$ is the pre-seeing shear polarizability tensor.
Since the measurement of $P^g_{\alpha\beta}$ is very noisy for faint
individual galaxies, 
$P^g_{\alpha\beta}$ is calibrated using galaxies detected with high signal-to-noise ratio $\nu>30$, 
following \cite{Okabe14a}. Similar calibration procedures are developed by
\cite{Umetsu10} and \cite{Oguri12}. 
Our shear calibration on $g_\alpha$ results in $2$--$3\%$ accuracy \citep{Okabe14a}.

We measure magnitudes of galaxies using SExtractor \citep{SExtractor}.
The SExtractor configuration is optimized for shape
measurements of faint galaxies. 
For each object, we compute a total Kron-like magnitude and an aperture
magnitude for color measurements in the AB-magnitude system.
To perform color measurements,  the $i'$-band data are degraded to the
PSF of $g'$-band data. 
For aperture photometry, the aperture diameter is set to $1.5$ times
the worst-seeing FWHM. 
Finally, we combine SExtractor-based photometry with weak-lensing shape
measurements to create a weak-lensing-matched photometry catalog.

\subsection{Background Selection}

It is of prime importance for an accurate mass measurement to securely select background galaxies.
Contamination of unlensed member and/or foreground galaxies in the shear catalog leads to 
a systematic underestimation of cluster lensing mass, referred to as a dilution effect 
\citep[e.g.][]{Broadhurst05,Umetsu08,Umetsu10,Okabe10b,Okabe11,Okabe13,Okabe14a}.
The dilution effect is increasing as a projected cluster-centric radius is decreasing,
because the fraction of member galaxies to background galaxies is increasing.
The effect is more dominant for more massive clusters.
We select background galaxies in color-magnitude plane, following 
studies for clusters at $z\sim0.2$ or higher redshift \citep{Broadhurst05,Okabe10b,Okabe13}.
A majority of unlensed galaxies to dilute lensing signals are faint and small member galaxies 
around the red sequence and apparent magnitudes $i'\sim 22-26$ ABmag in the color-magnitude plane.
Although the red sequence of bright member galaxies are clearly identified in color-magnitude plane 
because of a passive evolution, 
the color distribution for faint galaxies becomes unclear as the apparent magnitude becomes fainter.
It is therefore very difficult to discriminate between member and background galaxies in the faint end.
We thus investigate the color distribution of member galaxies using lensing signals in order to minimize the contamination.
We measure a mean tangential distortion strength in the central region of clusters 
by changing lower limits of color for the background selection,
$\langle g_+ \rangle(>\Delta C) = \sum_i g_{+,i}(>\Delta C) w_{g,i}(>\Delta C)  / \sum_i w_{g,i}(>\Delta C)$.
Here, $\Delta C=(g'-i')-(g'-i')_{\rm E/S0}$ is as a function of color offset from the red sequence, 
$g_{+,i}$ is the tangential component of reduced shear of the $i$-th galaxies 
with respect to the cluster center, given by 
\begin{eqnarray}
g_{+,i}&=&-(g_{1,i}\cos2\varphi+g_{2,i}\sin2\varphi), \nonumber
\end{eqnarray}
and $w_{g,i}=1/(\sigma_{g,i}^2+\alpha^2)$ is a statistical weight.
Here, $\varphi$ is the position angle between the first coordinate axis on the sky and the
vector connecting the cluster center and the galaxy position. 
The position of the BCGs is adopted as the cluster center.
$\sigma_{g,i}$ is an rms error of the shear estimate and $\alpha$ is the softening constant variance 
representing the scatter due to the intrinsic ellipticity of the galaxies 
\citep[e.g.,][]{Hoekstra00,Hamana03,Okabe10b,Umetsu10,Oguri12}. We choose $\alpha=0.4$.

In order to clarify the dilution effect in noisy lensing signals, 
the mean distortion strength is cumulatively 
computed as a function of color offset from the red sequence, $\Delta C$.
We choose the central region of $100-500h_{70}^{-1}$~kpc from brightest cluster galaxies (BCGs).
We also compute the mean lensing depth $\langle D_{ls}/D_s\rangle$.
Since it is difficult to estimate photometric redshifts of individual objects using two bands,
we used the COSMOS photometric redshift catalog \citep{Ilbert09} estimated by combining 30 bands.
The mean photometric redshift is computed with a statistical weight of $w_g$ 
by matching with the COSMOS photometric redshift in color-magnitude plane.
If there were no contamination of unlensed member galaxies,
it is expected that a ``Dilution'' estimator, which is defined by the ratio between the mean lensing signal and depth,
 $D=\langle g+\rangle/\langle D_{ls}/D_s\rangle(>\Delta C)$, is constant irrespective of the lower limit of the color offset, $\Delta C$.
Here, $D_{ls}$ and $D_s$ are the angular diameter distance between cluster (lens) and source redshifts and between observers and source redshifts, respectively.
Therefore, $D$ is a good estimator to investigate the color distribution of member galaxies.

\begin{figure}
\includegraphics[width=\hsize]{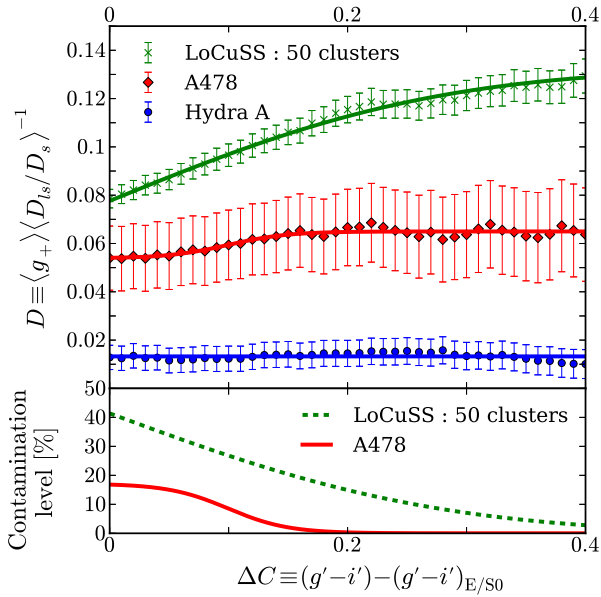}
\caption{Top: dilution estimator,
$D=\langle g+\rangle/\langle D_{ls}/D_s\rangle$, 
as a function of color offset from the cluster red sequence. 
The red diamonds, blue circles, and green crosses show $D$ for A478, Hydra
A, and an ensemble of 50 clusters from the LoCuSS project, 
respectively. 
The relative lensing strengths $D$ for A478 and the LoCuSS 50 clusters
 increase with increasing $\Delta C$, while $D$ for Hydra A 
is randomly distributed. 
The solid lines represent the best-fit models to describe the color
 distribution of member galaxies. 
Bottom: contamination level calculated by the respective 
best-fit model as a function color offset. 
The red-solid and green-dashed lines show the results of A478 and the
 LoCuSS 50 clusters, respectively.}
\label{fig:Dilution}
\end{figure}

The resultant $D$ is shown in the top panel of Figure \ref{fig:Dilution}.
The red diamonds and blue circles denote $D$ for A478 and Hydra A, respectively.
Although the error is large, $D$ of A478 is constant at $\Delta C\sim0$, 
increases with $\Delta C$ at $\Delta C\sim0.1$, and becomes flat at $\Delta C>0.3$.
The small $D$ at $\Delta C\sim0$ indicates a presence of member galaxies.
On the other hand, $D$ of Hydra-A shows a random distribution for $\Delta C$, 
suggesting no significant feature of the contamination.
We model a color distribution of $D$, given by $A+B\tanh((\Delta C-\Delta C_0)/\sigma)$.
Here, $A$ is the normalization at $\Delta C=0$, $A+B$ is a mean lensing strength of pure background galaxies,
and the second term represents a constant $D$ at $\Delta C\sim0$ and $\Delta C >0.3$.
As shown in Figure \ref{fig:Dilution}, the best-fit model (red solid line) of A478 well describes the data.
As for Hydra A, the best-fit model is consistent with no contamination.
We derive a color distribution of contamination level of unlensed
galaxies in background shear catalog (the bottom panel of Figure \ref{fig:Dilution}).
The maximum contamination level is at most $17\%$ at $\Delta C=0$.
In order to investigate a cluster-redshift dependence of the dilution effect,
we compute the mean $D$ within the same physical radius for 50 clusters at $z\sim0.2$
from the Local Cluster Substructure Survey (LoCuSS).
Here, $\Delta C$ is the color offset from the red sequence in the $i'-V$ band. 
The calculation and model for stacked 50 clusters are described in \cite{Okabe13}.
The maximum contamination level for 50 clusters (green dashed line in
the bottom panel) is $\sim 40\%$ at $\Delta C=0$.
The virial mass of A478 is more massive than the average of 50 clusters
(Section \ref{sec:WLmass}) and our data is shallower than those of 50 clusters, 
but nonetheless the contamination is less than half of those of clusters $z\sim0.2$.

There are two reasons for low contamination of the very nearby cluster ($z\sim0.09$).
First, since the colors of red-sequence galaxies become more blue as the redshift decreases,  
the number of red background galaxies increases.
Second, the ratio of member to background galaxies for very nearby
clusters becomes lower because an apparent covering area becomes larger.
Therefore, the dilution effect for clusters at lower redshifts becomes drastically less significant.
It is consistent with the feature of the very nearby cluster of Hydra-A ($z\sim0.05$) .

In order to securely select background galaxies, we employ $1\%$ contamination level following \cite{Okabe13} 
and define background galaxies with $\Delta C>0.17$ for A478. 
As for Hydra A, we do not apply the color cut because of the tiny contamination level,
because $D$ is almost constant in the wide color range of $-1<\Delta C<1$.
In order to assess our background selection for Hydra A, we make another background catalog 
selected with the same color cut of A478, conduct a weak lensing mass measurement (Section \ref{sec:WLmass})
and confirm the result dose not change.
The background number densities for A478 and Hydra A are $n_{\rm bkg}=6.2\,[\rm arcmin^{-2}]$ and
$28.9\,[\rm arcmin^{-2}]$, respectively.
The mean source redshifts for A478 and Hydra A are $\langle z_s\rangle\simeq0.643$ and 
$\langle z_s\rangle\simeq0.591$, respectively.

\section{Mass Measurements} \label{sec:WLmass}

Tangential distortion signals 
contain full information on the lensing signals of clusters.
Since clusters can be considered as gravitationally bound spherical objects at zeroth order,
a measurement of tangential shear profile is a powerful way to measure cluster masses.
The tangential shear is measurable by averaging over sufficient background galaxies, 
in order to suppress shape noise attributable to random intrinsic ellipticities of background galaxies.
We measure a mean tangential shear $\langle g_{+,i}\rangle$ 
as an average of the tangential shear component of background galaxies residing in the $i$-th radial bin 
with the statistical weight of $w_g$.
The other distortion component, $g_\times$, which is the $45$ degree rotated component, 
is also estimated. Hereafter, we refer to $\langle g_{+(\times),i}\rangle$ as $g_{+(\times),i}$.

In order to interpret the lensing signals, we here briefly
introduce mass models as the cluster mass density profile.
The reduced tangential shear profile for mass models can be expressed by
\begin{eqnarray}
 g_+(\theta)=\frac{\bar{\kappa}(<\theta)-\kappa(\theta)}{1-\kappa(\theta)},\label{eq:g+}
\end{eqnarray}
where $\bar{\kappa}$ and $\kappa$ are the mean surface mass density
within a radius $\theta$ and the surface mass density at $\theta$, respectively.
The lensing signals are computed by $\bar{\kappa}$ and $\kappa$ for
given mass models.

The NFW model \citep{NFW96,NFW97} is a theoretically well-motivated mass model
according to numerical simulations of dark matter particles. 
They found that the matter density profile of CDM halos is well described by 
an analytic function 
\begin{equation}
\rho_{\rm NFW}(r)=\frac{\rho_s}{(r/r_s)(1+r/r_s)^2},
\label{eq:rho_nfw}
\end{equation}
where $\rho_s$ is the central density parameter and $r_s$ is the scale radius.
The asymptotic mass density slopes are $\rho\propto r^{-1}$ and $r^{-3}$ for $r\ll r_s$ and $r\gg r_s$, respectively.  
The NFW profile is specified by two parameters of $M_\Delta$ 
which is a mass enclosed within a sphere of radius $r_\Delta$,
and the halo concentration $c_\Delta=r_\Delta/r_s$.
Here, $r_\Delta$ is the radius inside of which the mean density 
is $\Delta$ times the critical mass density, 
$\rho_{\rm cr}(z)$, at the redshift, $z$. 
For a comparison, we model a singular isothermal sphere (SIS) profile for
the main halo, given by 
\begin{equation}
\rho_{\rm SIS}(r)=\frac{\sigma_v^2}{2\pi G r^2},
\label{eq:rho_sis}
\end{equation}
where $\sigma_v^2$ is a one-dimensional velocity dispersion.
We also consider the BCG mass model to estimate the lensing signals for the BCGs.
Since the innermost radii for the profiles are outside the BCGs, 
the tangential shear from the BCGs is sensitive to the interior
mass within $\theta$, in other words, $\kappa(\theta)=0$ in Equation (\ref{eq:g+}).
Since the internal structure for the BCGs cannot be resolved by lensing
signals, 
we reasonably employ the point mass, $M_{\rm pt}$, 
to estimate the BCG mass, in terms of $g_+(\theta)\propto M_{\rm pt}/(\pi \theta^2)$.

Given specific mass models, we perform $\chi^2$ fit to the tangential shear profile, $g_+$.
The $\chi^2$ is defined as followed: 
\begin{eqnarray}
  \chi_g^2&=&\sum_{i,j} \left( g_{+,i}  - g_{+,i}^{\rm model}\right)C_{ij}^{-1}\left( g_{+,j} - g_{+,j}^{\rm model}\right),
\end{eqnarray}
where 
\begin{eqnarray}
C_{ij}=C_{{\rm stat},ij}+C_{{\rm LSS},ij}.  \label{eq:Cnoise}
\end{eqnarray}
The first term of $C$ is the statistical noise relevant to 
the intrinsic ellipticity of background galaxies.
The statistical noise for the shear is given by $C_{{\rm stat},ij}=\bar{\sigma}^2_{g,i}\delta_{ij}/2$,
where $\bar{\sigma}_{g,i}$ is the statistical uncertainty of the two components in the $i$-th radial bin 
and $\delta_{ij}$ is Kronecker's delta. 
The second term, $C_{{\rm LSS},ij}$, 
is the error covariance matrix of uncorrelated large-scale structure (LSS) 
along the line-of-sight, \citep[e.g.][]{Schneider98,Hoekstra03,Umetsu11,Umetsu14,Oguri10b,Oguri11a,Okabe13,Okabe14a}, 
estimated from the weak-lensing power spectrum with WMAP7 cosmology \citep{WMAP07}. 
The cosmic shear correlation function leads to non-zero off-diagonal elements for $C_{{\rm LSS},ij}$.

The signal-to-noise ratio (S/N) for the tangential shear profile is computed by
\begin{eqnarray}
 \left(\frac{S}{N}\right)^2= \sum_{ij} g_{+,i}C_{ij}^{-1}g_{+,j}.
\end{eqnarray}

\subsection{A478}\label{subsec:a478}

\begin{figure}
\includegraphics[width=\hsize]{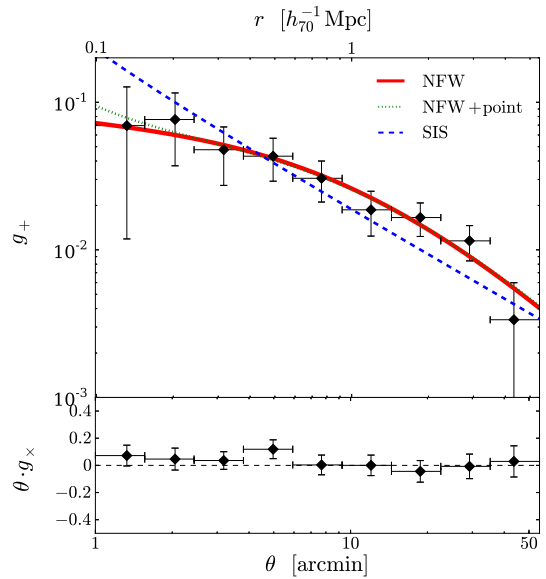}
\caption{Radial profiles of the tangential reduced-shear component (top panel),
 $g_+$, and the $45^\circ$-rotated component (bottom panel),
 $\theta\cdot g_\times$, for A478. 
The vertical error bar of each data point ($i$-th bin) shows the diagonal
 element of the error covariance matrix, $C_{ii}^{1/2}$ (Equation
 \ref{eq:Cnoise}),
 for the top panel and
 $\theta_i C_{ii}^{1/2}$ for the bottom panel, respectively. 
The tangential reduced-shear profile shows a clear radial curvature, in
 agreement with the steepening NFW profile.
The red-solid, green-dotted, and blue-dashed lines show the best-fit NFW,
 NFW+point, and SIS profiles, respectively. 
}

\label{fig:g+_a478}
\end{figure}

The tangential reduced-shear profile for A478 is shown in Figure \ref{fig:g+_a478},
in a range of $100{\rm kpc} \simlt r \simlt 5.5{\rm Mpc}$.
The innermost radius, determined by requiring a sufficient number of
background galaxies, is larger than the effective radius of the BCG.
The S/N of the tangential shear profile is $8.1$.
The tangential shear profile exhibits a clear curvature.
On the other hand, the B-mode, $g_\times$, shows both negative and
positive values, which is consistent with null signal.
We plot $\theta \cdot g_\times(\theta)$ in the bottom panel of Figure
\ref{fig:g+_hydraa}, so that the scatter is independent of radius
for logarithmically spaced binning \citep{Miyatake13}.

We fit a single NFW model or two components of the NFW model as the
smooth matter density profile and the
point source ($g_+=g_{\rm NFW}+g_{\rm pt}$ ; NFW+point) 
to the tangential shear profile (Table \ref{tab:mass}).
The minimum chi-square are $\chi^2_{\rm min}({\rm d.o.f})=2.18(7)$ 
and $2.12(6)$ for the first and second model, respectively, 
where the degree of freedom is presented in the round bracket.
Here, since errors are mainly attributable to the variance of intrinsic
ellipticity, $\langle \sigma_g^2\rangle^{1/2}\sim0.4$, in addition to measurement errors, $\chi^2_{\rm min}$ becomes small.
It shows that two models are acceptable as the cluster mass model. 
In order to estimate the validity of the additional point source, 
we calculated F test probability, $0.7$, indicating that there is no strong reason to 
add the point source to describe the current data.
The best-fit SIS profile ($\sigma_v= 958.67_{-62.96}^{+58.11}~{\rm
kms^{-1}}$) is 
deviated from the curvature of the tangential shear profile
, albeit still acceptable 
($\chi^2_{\rm min}({\rm d.o.f})=9.13(8)$).

\subsection{Hydra A} \label{subsec:hydraa}

\begin{figure}
\includegraphics[width=\hsize]{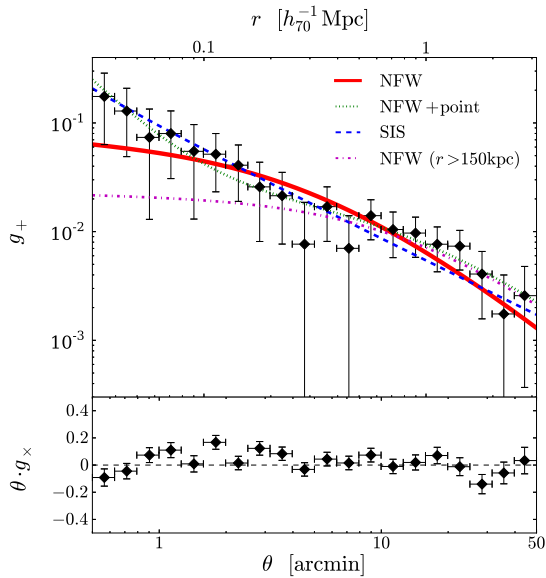}
\caption{Radial profiles of the tangential reduced-shear component (top panel),
 $g_+$, and the $45^\circ$-rotated component (bottom panel),
 $\theta\cdot g_\times$, for the Hydra A  cluster, shown over a wide 
 radial range of $30\mathrm{kpc}\simlt r\simlt3\mathrm{Mpc}$. 
The red-solid, magenta-dot-dashed, green-dotted, and blue-dashed lines 
show the best-fit NFW profiles derived from the data at $r>30$\,kpc and $r>150$\,kpc, 
the best-fit NFW+point and SIS profiles, respectively.
}
\label{fig:g+_hydraa}
\end{figure}

A large apparent size of Hydra A allows us to estimate the tangential shear profile 
in a wide radial range of $30~{\rm kpc}\simlt r\simlt3~{\rm Mpc}$, as shown in Figure \ref{fig:g+_hydraa}.
In particular, lensing signals at $r\simlt 50-100~{\rm kpc}$ is detected solely by weak-lensing analysis.
It is an advantage of weak-lensing study of very nearby clusters,
because cluster central regions at higher redshifts are in the strong-lensing regime.
For instance, an innermost radius of a stacked lensing profile for 50 clusters 
$z\sim0.2$ \citep{Okabe13} is $70 h_{70}^{-1}$kpc.
The tangential profile is highly detected at $\mathrm{S/N}=7.0$.
The profile shows two characteristic features : 
a sharp decrease in the central region ($r<100~{\rm kpc}$)
and a clear curvature in the range of $100~{\rm kpc} \simlt r \simlt 3~{\rm Mpc}$.
The B-mode values, $g_\times$, are randomly distributed cross the null signal.

First, we fit the profile with the NFW model 
and obtain a high concentration parameter $c_{\rm vir}=9.37_{-3.94}^{+6.99}$ with virial mass 
$M_{\rm vir}=2.74_{-1.21}^{+1.99}\times10^{14}\Msol$.
The best-fit mass profile shows a clear curvature (Figure \ref{fig:g+_hydraa}).
However, we found that the best-fit model is lower than 
the observed lensing signals at $r\simlt50~{\rm kpc}$ and $r\simgt2~{\rm Mpc}$.
It suggests that the virial mass is underestimated
because the mass estimate is sensitive to lensing signals in the outskirts.
A similar deviation 
is found in the best-fit SIS model ($\sigma_v=631.93_{-47.77}^{+43.71}\,{\rm kms^{-1}}$).
On the other hand, when we fit the NFW model to the lensing profile beyond $150~{\rm kpc}$, 
the best-fit concentration becomes lower $c_{\rm vir}=4.52_{-1.75}^{+2.95}$ (Table \ref{tab:mass}).
The best-fit profile well describes the data at $r\simgt300~{\rm kpc}$, 
but significantly underestimates the signals at $r\simlt100~{\rm kpc}$.

It indicates that the inner slope of the tangential profile 
becomes stepper than expected from the outskirts.
The steep inner slope suggests
one possibility that an extra component exists inside the innermost radius.
In particular, the slope for the lensing profile at a few inner bins is proportional to $\sim-2$.
We thus consider two components of the NFW model and the point mass (NFW+point).
The best-fit parameters are shown in Table \ref{tab:mass}.
The $\chi^2_{\rm min}({\rm d.o.f})$ is improved from $8.62(18)$ to $4.02(17)$.
The best-fit profile well describes the data in the full radial range (Figures \ref{fig:g+_hydraa}).
In order to estimate the validity of the additional point source, 
we conducted F test and found that the probability $4\times10^{-5}$ is significantly small. 
It is thus reasonable to add the extra component to the smooth mass
component for describing the observed lensing signals.
In other words, the two-components model better describes the tangential profile.



\begin{table*}
\caption{Mass estimates for Hydra A and A478. 
The NFW mass ($M_\Delta^{\rm NFW}$) and the point mass ($M_{\rm pt}$)
 are in units of $10^{14}\h70Msol$ and $10^{12}\h70Msol$,
 respectively. 
} \label{tab:mass}
\begin{center}
\begin{tabular}{lccccccc}
\hline
Cluster           & model \& method
                  & $M_{\rm vir}^{\rm NFW}$ 
                 & $M_{\rm 200}^{\rm NFW}$ 
                 & $M_{\rm 500}^{\rm NFW}$  
                 & $M_{\rm 1000}^{\rm NFW}$
                 & $M_{\rm 2500}^{\rm NFW}$
                 & $M_{\rm pt}$\\
\hline
A478        &  NFW
            &  $16.38_{-4.41}^{+5.70}$
            &  $12.93_{-3.22}^{+3.96}$
            &  $8.67_{-1.88}^{+2.14}$
            &  $5.95_{-1.17}^{+1.28}$
            &  $3.17_{-0.72}^{+0.72}$
            &  $-$\\
A478        &  NFW+point
            & $16.72_{-4.65}^{+6.30}$
            & $13.05_{-3.30}^{+4.12}$
            & $8.56_{-1.97}^{+2.19}$
            & $5.74_{-1.53}^{+1.55}$
            & $2.93_{-1.23}^{+1.28}$
            & $<25.90$\\
Hydra A       &  NFW ($r>150~h_{70}^{-1}{\rm kpc}$)
              &  $4.66_{-1.82}^{+2.77}$
              &  $3.62_{-1.31}^{+1.85}$
              &  $2.40_{-0.76}^{+0.95}$
              & $1.63_{-0.46}^{+0.55}$
              &  $0.85_{-0.28}^{+0.29}$
              & $-$ \\ 
Hydra A       & NFW+point (full radial range)
              & $4.88_{-2.04}^{+3.29}$
              & $3.66_{-1.41}^{+2.05}$
              & $2.27_{-0.77}^{+0.98}$
              & $1.43_{-0.48}^{+0.55}$
              & $0.64_{-0.31}^{+0.32}$
              & $6.10_{-2.50}^{+2.47}$ \\
\hline
\end{tabular}
\end{center} 
\end{table*}

\section{Joint Analyses} \label{sec:dis}

\subsection{Joint Weak-lensing and Stellar Photometry Study of A478} \label{subsec:joint_a478}

We study the mass distribution of A478 using weak-lensing and
BCG (PGC014685) photometry measurements.

The tangential shear profile (Section \ref{subsec:a478}) for
A478 is well described by a single NFW model or
a two-component mass model including the BCG as an unresolved point
mass. 
Complementary information at small scales is useful to directly
determine the BCG mass profile as well as to better constrain the total
mass profile in combination with lensing data.

For this purpose, we use stellar mass estimates of the BCG from a photometry.
We estimate the stellar mass for the BCG using $K$-band photometry
using the $K$-correction and galactic extinction from 2MASS extended
source catalog \citep{2MASS06}.  
Since the $K$-band luminosity is sensitive to the light of old stars and less
sensitive to recent star-formation activities,  it serves as a
reasonable proxy for stellar mass.
The difference in $K$-band transmission between the 2MASS and traditional
Johnson $K$ filters is not important for our purpose \citep{Carpenter01}.
We employ both 
empirical \citep{Arnouts07} and theoretical \citep{Longhetti09} scaling relations 
between the stellar mass and $K$-band luminosity.
We consider the Salpeter \citep{Salpeter55}, Chabrier \citep{Chabrier03}
and Kroupa \citep{Kroupa01} IMFs.
The mass uncertainty is estimated by taking into account the
 photometric errors, $15\%$ intrinsic scatter for the
empirical scaling relation \citep{Arnouts07}, and systematic
uncertainties from different scaling relations. 

\begin{figure}
\includegraphics[width=\hsize]{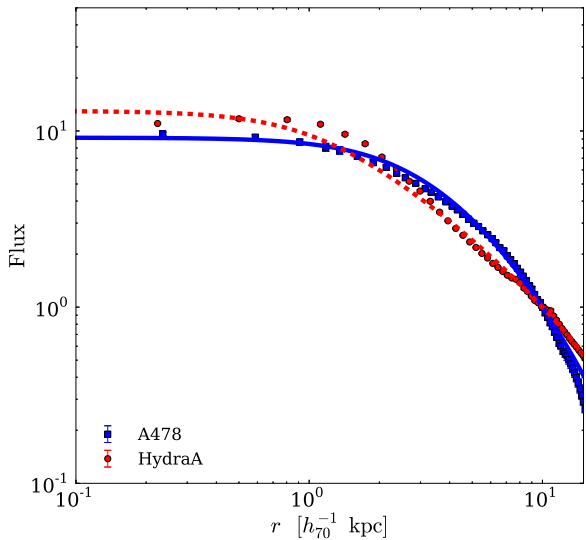}
\caption{Projected surface luminosity profiles for BCGs in A478 (red
 circles) and the Hydra A cluster (blue squares) in arbitrary units. 
The solid and dotted lines denote the best-fit profiles for A478 and the
 Hydra A cluster, respectively}
\label{fig:cDprof}
\end{figure}

We assume a pseudo isothermal mass distribution (PIEMD) model
\citep[][]{Elasdttir07,Limousin09,Natarajan09} to describe the stellar
mass distribution of the BCG.
The PIEMD model is specified with three parameters, namely,
the core radius ($a$),
the scale radius ($s$), 
and the normalization ($\rho_0$), in the following form:
\begin{eqnarray}
 \rho=\frac{\rho_0}{(1+r^2/a^2)(1+r^2/s^2)}~~~~(s\geq a). 
\label{eq:PIEMD}
\end{eqnarray}
The projected sufrace luminosity profile for the BCG is shown in
Figure \ref{fig:cDprof}. The profile is fitted with the PIEMD model 
convoluted with the PSF kernel. We find that the outer edge of the
luminosity profile is slightly truncated and deviated from expectations
for the PIEMD model. Although the truncated PIEMD model would be better to fit
the overall profile, the deviation is small for this study and thus we adopt
the PIEMD model here.
We obtain $a=2.7$kpc and $s=11.3$kpc.
Assuming a scaling relation
between cuspy radius and the luminosity for BCGs \citep{Lauer07}, the
mean core radius is expected to be $a\sim0.8$kpc. 
The best-fit core radius is larger than the mean one calculated from the scaling
relation, which might be
caused by a limitation of ground-based telescope because the apparent
size of the expected core radius, $\sim0.5$arcsec, is small. 
We find the best-fit solution by minimizing the combined $\chi^2$
function, $\chi^2=\chi^2_g+\chi^2_s$, with $\chi_s^2$ defined by 
\begin{eqnarray}
 \chi^2_s=\frac{(M_*-M_{\rm PIEMD,*})^2}{\delta_{M_*}^2}.
\end{eqnarray}
Here, $M_*$ and $\delta_{M_*}$ are the stellar mass and its
uncertainty estimated with specific IMFs within the measurement radius $17$\,kpc, respectively.
$M_{\rm PIEMD}$ is the model for the BCG.
We determine the normalization of the PIEMD model and the NFW parameters
from the joint fit.

\begin{figure}
\includegraphics[width=\hsize]{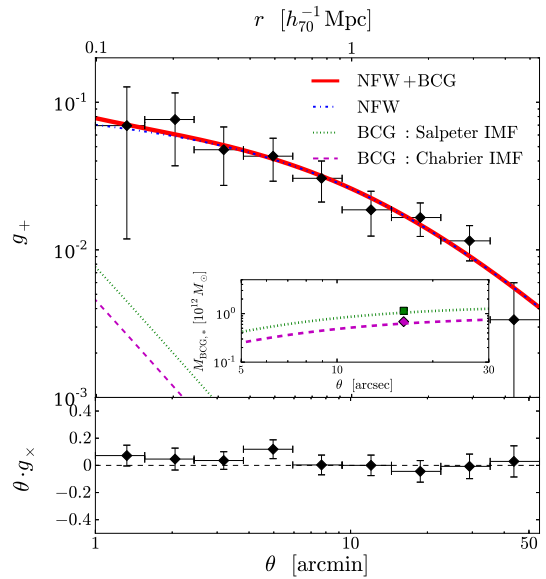}
\caption{Tangential reduced-shear profile and BCG stellar mass
 profile (inset) for A478.
The red-solid line represents the best-fit solution for the total lensing
 signal with the NFW (blue dot-dashed line) plus BCG (Salpeter IMF) model.
The green-dotted and magenta-dashed lines show the best-fit lensing 
 profiles and BCG stellar mass profiles, obtained with the Salpeter and Chabrier
 IMFs, respectively.}
\label{fig:WL+Mstell}
\end{figure}

Figure \ref{fig:WL+Mstell} shows the best-fit model for the tangential
shear profile and the BCG stellar mass profile.
The Chabrier and Kroupa IMFs give total stellar mass estimates of 
$M_{\mathrm{BCG},*}^\mathrm{Cha}\sim M_{\mathrm{BCG},*}^\mathrm{Kro}\sim 8\times10^{11}M_\odot$, 
compared to
$M_{\mathrm{BCG},*}^\mathrm{Sal}=(15\pm2)\times10^{11}M_\odot$ with the
Salpeter IMF,
where the errors mainly account for systematic uncertainties from
different scaling relations.
The IMF uncertainty gives rise to a systematic bias on stellar mass estimates,
as pointed out by \cite{Oguri14}.
Hence, independent constraints, such as from stellar kinematics 
\citep{Newman09,Newman11,Newman13}
quasar microlensing
\citep{Oguri14} observations, are needed to meaningfully 
break the IMF uncertainty (see Section \ref{subsec:joint_hydraA}). 
Despite the IMF uncertainty, 
the best-fit NFW mass model 
($M_\mathrm{vir}=16.48\times10^{14}\Msol$ and
 $c_\mathrm{vir}=4.56_{-1.28}^{+1.67}$) 
is not sensitive to the choice of IMF because the lensing signal 
from the main cluster halo is sufficiently high.
We note that the result does not significantly change when we adopt $a=0.8$kpc.

\subsection{Joint Weak-lensing and Stellar Kinematics Study of Hydra A}
\label{subsec:joint_hydraA}

The tangential shear profile for the Hydra A cluster 
is well described by the two-component mass model of the BCG's point mass
and the smooth NFW profile (Section \ref{subsec:hydraa}).
The smooth mass component is determined by fitting lensing signals at
outer cluster radii, whereas the point mass measurement is sensitive to
lensing signals at innermost bins because the BCG lensing signal is embedded
in the lensing signals of the main cluster.
Recently \citet{Newman09,Newman11,Newman13} 
performed a joint analysis of
weak/strong-lensing and stellar kinematics data to break the IMF uncertainty and 
measure the mass profile from the BCG to the cluster main halo.
They demonstrated that the BCG stellar kinematics is powerful for
determining the BCG mass profile and the cluster inner mass profile 
under two assumptions of dynamic equilibrium and mass tracing light.
Following their pioneering studies, 
we here conduct a joint weak-lensing and stellar-kinematics analysis of
Hydra A. We note that there are no publicly available data of stellar
kinematics for A478.
Since the Hydra A cluster is at a very low redshift, strong-lensing
information is not required for our joint analysis.

We here summarize the dynamical properties of the BCG (3C218).
It is well known that the AGN in the BCG ejects a pair of jets, showing interaction 
between the jets and surrounding ICM \citep{McNamara00}.
\cite{Fujita13} 
have discovered a dust lane in the central region of the BCG ($4~{\rm kpc}\times0.8~{\rm kpc}$).
The major axis is along the east-west direction, nearly perpendicular to the jet axis.
\cite{Hamer14} found that the dust lane is a rotating disk by measuring
several line emissions. 
The maximum rotating velocity reaches $\sim400$\,km\,s$^{-1}$. 
\cite{Melnick97} also found a fast rotating disk using the H$\alpha$
line.
\cite{Ekers83} measured stellar velocities 
and pointed out that the BCG is composed of 
a rotating disk of stars embedded in non-rotating stellar components.
The rotational velocity inside $4\arcsec$ ($\sim 4$\,kpc) is found to be
$\sim 50$\,km\,s$^{-1}$ \citep{Loubser08}.
The stellar velocity dispersion is $\sim 200$--$300$\,km\,s$^{-1}$
\citep{Heckman85,Loubser08}.
Therefore, the central part of the BCG ($\simlt 4\arcsec$) 
comprises cold rotating disk, rotating stellar, and non-rotating stellar
components, exhibiting complex dynamics.
On the other hand, the non-rotating stellar components dominate 
outside the rotating disk ($\simgt4\arcsec$).

\begin{figure*}
\includegraphics[width=0.45\hsize]{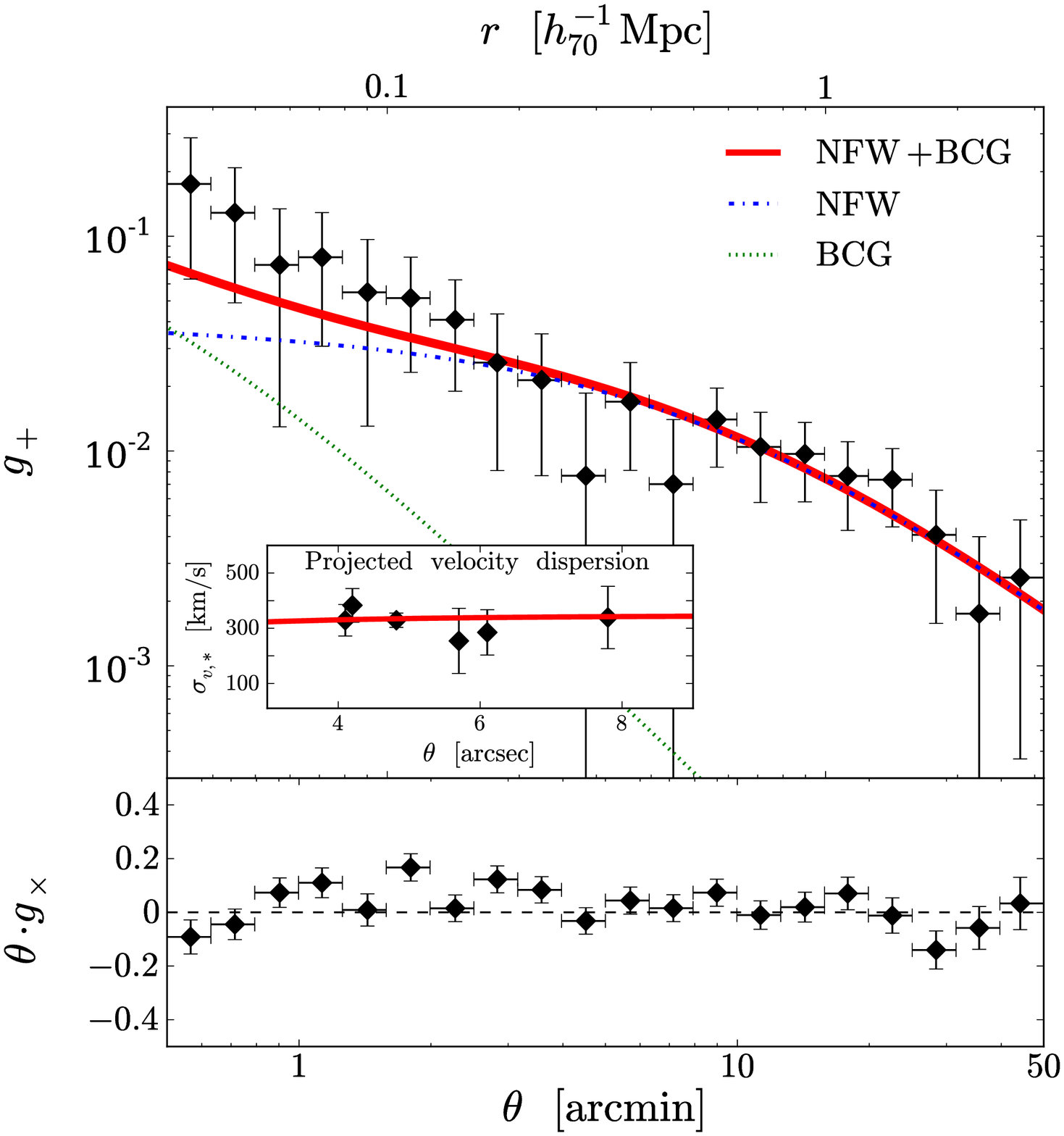}
\includegraphics[width=0.45\hsize]{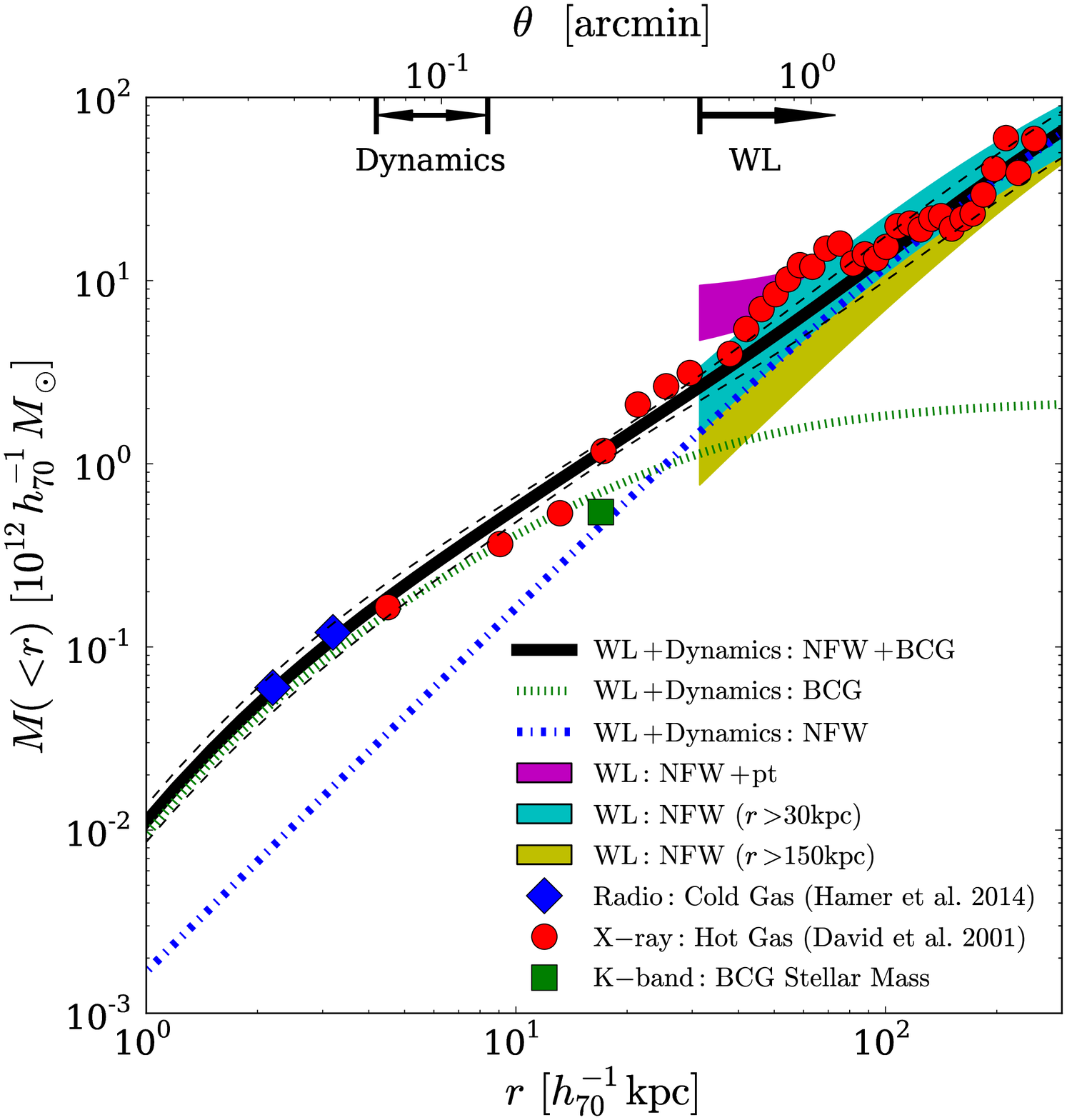}
\caption{Left: tangential reduced-shear for the Hydra A cluster and
 projected velocity dispersion for the central BCG (inset). 
Circles and triangles denote \citet{Heckman85} and
 \citet{Loubser08} for projected velocity dispersion, respectively.
The red-solid lines represent the best-fit models for the total lensing
 signal and the projected velocity dispersion.
The dotted-green and blue-dot-dashed lines show the lensing signals for
 the BCG and NFW mass components, respectively.
Right: spherical mass profile within $300$\,kpc.
The back-solid and dashed-lines represent the total mass profile and its
 $1\sigma$ uncertainty, respectively.
The green-dotted and blue-dot-dashed lines represent the BCG stellar and
 NFW halo components, respectively.
The magenta, cyan, and yellow shaded areas (from the top to the bottom) show
 the $1\sigma$ error regions for the 
NFW+point,
NFW ($r>30$\,kpc),
and NFW ($r>150$\,kpc) 
models from the weak-lensing-only fits, respectively.
The blue diamonds, red circles, and green squares show dynamical mass
 estimates from radio-line measurements of cold gas \citep{Hamer14},
 hydrostatic mass estimates from X-ray emitting hot gas \citep{David01}, 
and stellar mass estimates from $K$-band photometry assuming the Salpeter
 IMF, respectively. 
The top arrows indicate the radial ranges for the dynamical
 data and weak-lensing data.}
\label{fig:WL+dynamics}
\end{figure*}

We use the PIEMD model (Equation \ref{eq:PIEMD}),
assuming spherically symmetry, isotropic orbits, and mass tracing light.
We determine the core and scale radii by fitting the light distribution
(Figure \ref{fig:cDprof})
and then estimate the normalization of the PIEMD and the NFW parameters
by a joint fit to weak-lensing and stellar velocity dispersion
measurements.
The combined $\chi^2$ function is defined by
$\chi^2=\chi_\mathrm{g}^2+\chi^2_\mathrm{dyn}$, where
\begin{eqnarray}
\chi^2_{\rm dyn}=\sum_i \frac{(\sigma_{v,*}-\sigma_{\rm l.o.s})^2}{\delta_{\sigma_{v*}}^2}
\end{eqnarray}
with $\sigma_{v,*}$ and $\delta_{\sigma_{v*}}$ 
the velocity dispersion and its uncertainty, respectively.
The velocity dispersion along the line-of-sight, $\sigma_{\rm l.o.s}$, 
is computed by the spherical Jeans equation \citep[eq. 42 of][]{Cappellari08}
\begin{eqnarray}
  \Sigma_* \sigma_{\rm l.o.s}^2(R)=2 G \int^\infty_R
   \frac{\nu_*(r)M(r)F(r)}{r^{2-2\beta}}dr \label{eq:dynamics} 
\end{eqnarray}
\citep[see also eq 12 of ][]{Newman13},
where 
$M(r)=M_{\rm PIEMD,*}(r)+M_{\rm NFW}(r)$ is
the total mass, 
$\beta=1-\sigma_\theta^2/\sigma_r^2$ presents the velocity anisotropy,
and $\nu_*$ and $\Sigma_*$ are the spherical and surface density
profiles of the stellar components, respectively; here we use the
best-fit light distribution of the PIEMD model to describe $\nu_*$ and $\Sigma_*$.
The kernel $F(r)$ is $(r^2-R^2)^{1/2}$ for the isotropic-orbit case,
$\beta=0$ \citep[see][]{Cappellari08,Elasdttir07}.
\cite{Lemze12} studied the radial dependence of $\beta(r)$ for simulated
cluster-sized dark-matter halos, finding 
$\beta\sim 0-0.2$ at $\sim 0.05 r_{\rm vir}$. 
Since the line-of-sight integral is weighted by the stellar density and
thus dominated by the innermost region, the results are insensitive to
$\beta(r)$ around the BCG. 
Hence, we can reasonably assume a constant $\beta$.
We fix $\beta=0$ throughout the paper. 
We also checked that these results are not sensitive to the choice of
$\beta$
and do not change significantly when fitting with $\beta=0.2$.
We use the stellar velocity dispersion at $r>4\arcsec$ where the
contribution from the non-rotating component is negligible.
We summarize the lensing data, projected velocity dispersion and
the light profile for the BCG in Table \ref{tab:hydraAdata}.
The best-fit radial profiles of weak-lensing and projected velocity dispersion 
are shown in the left panel of Figure \ref{fig:WL+dynamics}.
The best-fit profile for the total lensing signal at $r\simlt 100$\,kpc
is systematically lower than the measurements, although consistent
within errors.

The spherical mass profiles $M(<r)$ in the central region 
($r<300$\,kpc) are shown in the right panel of Figure \ref{fig:WL+dynamics}.
The stellar mass component (green dotted line) dominates over 
the NFW halo component (blue dotted line) at $r\simlt20$\,kpc.
The NFW mass component from the joint fit is 
consistent with the best-fit NFW profile from the $g_+$ 
fitting using the data at $r>150$\,kpc (yellow region).
The total mass profile (black solid line) agrees 
with the weak-lensing-only results using the full-range data ($r>30$\,kpc; cyan region)
because these best-fit tangential profiles are similar at $r\simlt100$\,kpc 
(Figures \ref{fig:g+_hydraa} and \ref{fig:WL+dynamics}).
The NFW+point total mass profile from tangential-shear fitting 
(magenta region) is in good agreement at $r\simgt 150$\,kpc
with that from the joint analysis, but overestimates the
joint results at $r\simlt 50$\,kpc.
We shall discuss this point in the last paragraph of this section 
as well as in Section \ref{subsec:joint_hydraA2}.
We note that the best-fit BCG mass profile is extrapolated
beyond the maximum radius for the light-profile measurements 
($\sim30$\,kpc) assuming a constant mass-to-light ratio.
If the mass profile for the BCG is sharply truncated at this radius,
the BCG mass is $M_\mathrm{BCG}=1.0_{-0.4}^{+0.3}\times10^{12}M_\odot$.

For comparison, 
we overplot the dynamical masses estimated from
the cold gas disk component \citep[blue diamonds;][]{Hamer14} 
and hydrostatic masses from X-ray observations \citep[red
circles]{David01}.
A detailed comparison of weak-lensing and X-ray mass measurements at
mass overdensities $\Delta<2500$ are presented in \citet{Okabe14b}. 
\cite{Hamer14} assumed a rotating exponential disk for the central cold
disk and obtained dynamical mass estimates within $3.2$\,kpc and
$2.2$\,kpc using the H$\alpha$ and Pa$\alpha$ lines, respectively. 
\cite{David01} derived hydrostatic mass estimates using {\em Chandra}
data.

Overall, our results agree well with independent measurements
from dynamical masses
estimated from the cold gas disk and X-ray hydrostatic 
mass estimates, over a wide range of radii from a few kpc to 300\,kpc,
irrespective of different physical conditions and nature of cold gas
particles, hot diffuse plasmas, and stars used as mass tracers.
Although the effects of non-thermal pressure (e.g., sound waves, subsonic
gas motions, and cosmic rays) are expected to be
significant as triggered by AGN activities,
the X-ray and weak-lensing mass measurements in the central region are in
excellent agreement.
Since turbulent and/or rotational gas motions in the ICM can be directly
proved by the SXS onboard {\em Astro-H},
a complementary combination of kinetic information about the ICM, cold gas,
and stars and weak-lensing data will be able to provide an improved
understanding of the physical state of the cluster center.

We also estimate the stellar mass of the BCG using $K$-band luminosities
as described in Section \ref{subsec:joint_a478}.
An empirical scaling relation \citep{Arnouts07} gives $M_{{\rm
BCG},*}\sim5^{+3}_{-1}\times10^{11}M_\odot$,  
within a photometric aperture radius of $17$\,kpc.  
The errors represent a $15\%$ scatter in the stellar mass-to-light ratio
\citep{Arnouts07}. 
The inferred stellar mass is in remarkable agreement with the BCG
mass determined by our joint analysis (Figure \ref{fig:WL+dynamics}).
Using the theoretically-predicted scaling relation \citep[][Table
2]{Longhetti09}, we find
$M_{{\rm BCG},*}\sim4-6\times10^{11}M_\odot$ for the Salpeter IMF.
On the other hand, the Chabrier \citep{Chabrier03} and Kroupa
\citep{Kroupa01} IMFs give a stellar mass of
$\sim3\times10^{11}M_\odot$, which is slightly lower than our best-fit
estimate.

The best-fit tangential shear profile at $r\simlt 100$\,kpc
is somewhat discrepant with the data.
\cite{Newman09,Newman13} took into account a calibration factor, 
$m_\mathrm{WL}=g_{+,\mathrm{obs}}/g_{+,\mathrm{true}}$, 
for their lensing  model 
(i.e., $g_{+,\mathrm{model}}\rightarrow
m_\mathrm{WL} \times g_{+,\mathrm{model}}$), 
possibly caused by systematic shear calibration and/or
redshift errors,  
to solve the discrepancy between their lensing and dynamical data.
\cite{Newman09} found $m_{\rm WL}=0.80$, 
and \cite{Newman13} used a prior $m_{\rm WL}=0.89\pm0.05$,
meaning that their shear measurements are underestimated by $\sim
10\%$--$20\%$, relative to their dynamical measurements.

We repeated fitting including an additional correction factor
$m_{\rm WL}$ as a free parameter
and obtained $m_{\rm WL}=3.42\pm1.24$, which is extremely 
higher than our shear calibration uncertainty \citep[$2-3\%$;][]{Okabe13,Okabe14a,Okabe15c}.
It is therefore unlikely that the discrepancy is primarily caused by shear
calibration errors, suggesting alternative possibilities, such as the
choice of mass models.
We estimate lensing signals from member galaxies located in the
gap between the BCG-kinematics and weak-lensing data regions
($8\mathrm{kpc}\simlt r\simlt 30\mathrm{kpc}$), 
assuming a stellar mass-to-light ratio based on the Salpeter IMF. 
Including their contributions, the best-fit lensing profile is enhanced by $\sim20\%$,
which however is insufficient to account for the discrepancy.
In the joint analysis, we assumed that the BCG stellar mass traces
the light distribution, and did not explicitly include BCG's dark-matter
contribution relative to the NFW form.
If there is a significant and extended dark-matter contribution
associated with the BCG, the apparent discrepancy may be explained.
We shall examine this hypothesis in the next subsection.

\subsection{Additional Dark-Matter Component around the BCG in Hydra A} \label{subsec:joint_hydraA2}

For the Hydra A cluster, 
the total mass profile around the BCG has been tightly constrained by the
joint analysis of weak-lensing and dynamical data.
We find that the inferred stellar mass contribution 
is not sufficient to fully
explain the observed lensing signal 
$g_+\propto r^{-2}$ at $r\sim100$\,kpc.
This may imply that there is an additional dark-matter component
associated with the BCG. 
Here, we adopt as a working hypothesis that the central
dark-matter distribution $\rho_\mathrm{DM}(r)$ is shallower than that
of the stellar component $\rho_*(r)$, namely 
$\rho_\mathrm{DM}(r) < \rho_*(r)$ at $r\to 0$
and
$\rho_\mathrm{DM}(r) > \rho_*(r)$ at larger radii.
We have first attempted to fit with the generalized NFW model 
$\rho \propto(r/r_s)^{-\gamma}(1+r/r_s)^{-3+\gamma}$ 
and the cored NFW model 
$\rho \propto (1+r/r_c)^{-1}(1+r/r_s)^{-2}$, both described by three parameters,
to
account for the hypothetical dark-matter contribution.
We find both models fall short of the central tangential shear signal and 
fail to explain the discrepancy.

Now we employ the PIEMD model to describe the additional mass component
around the BCG.
We express the total mass in Equation (\ref{eq:dynamics}) as
$M(r)=M_{\rm PIEMD,*}(r)+M_{\rm PIEMD,DM}(r)+M_{\rm NFW}(r)$,
where $M_{\rm PIEMD,DM}(r)$ is specified by three parameters,
namely the normalization, core and scale radii.
We here assume the minimum core radius to be $0.4\arcmin$ 
in order to satisfy the requirement that the dark matter
distribution is shallower than the stellar distribution,  
because there is no visible tracer of hypothetical dark-matter profile, 
that is, the current data is not sufficient to constrain the shape of the profile.

\begin{figure*}
\includegraphics[width=0.45\hsize]{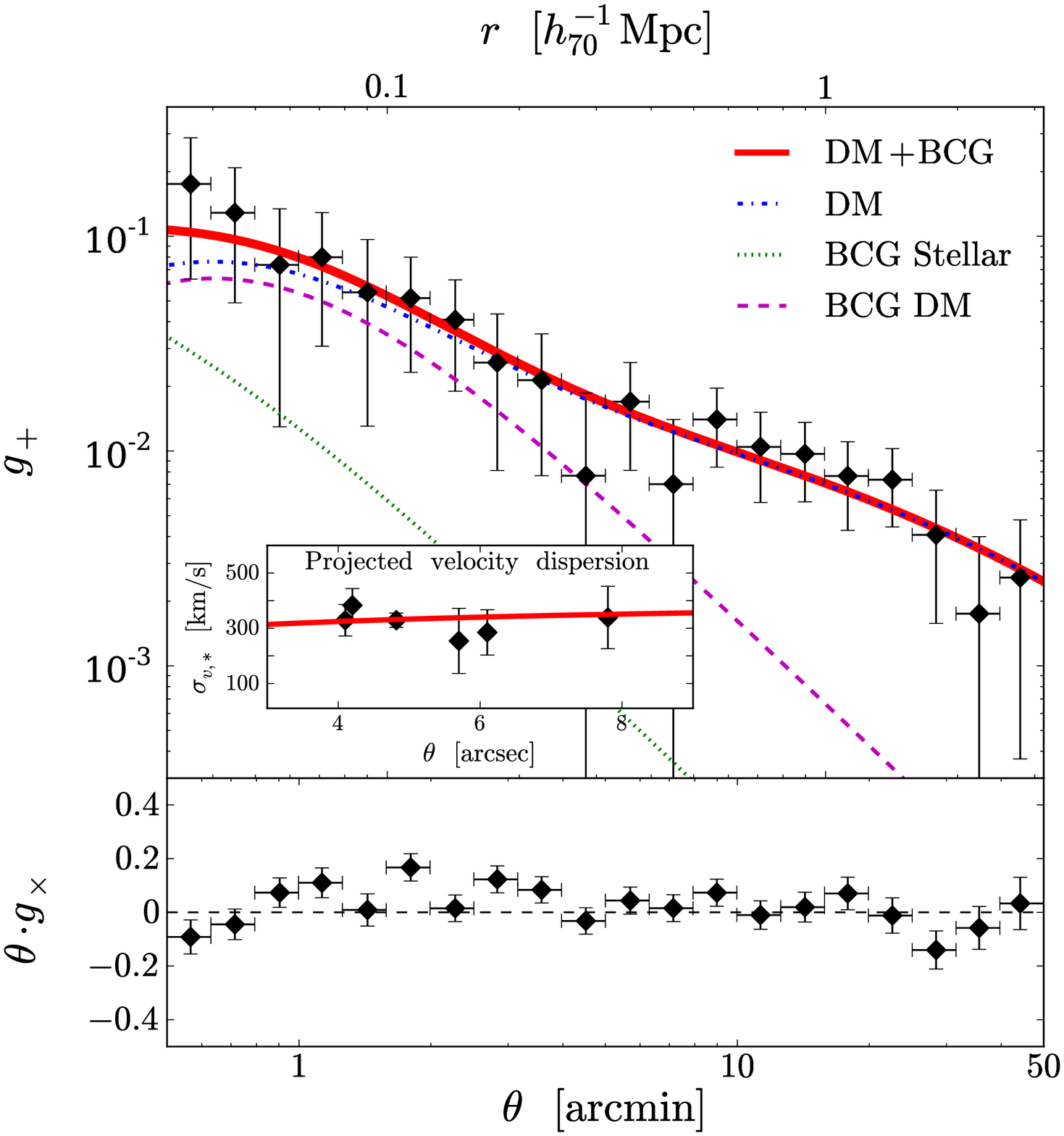}
\includegraphics[width=0.45\hsize]{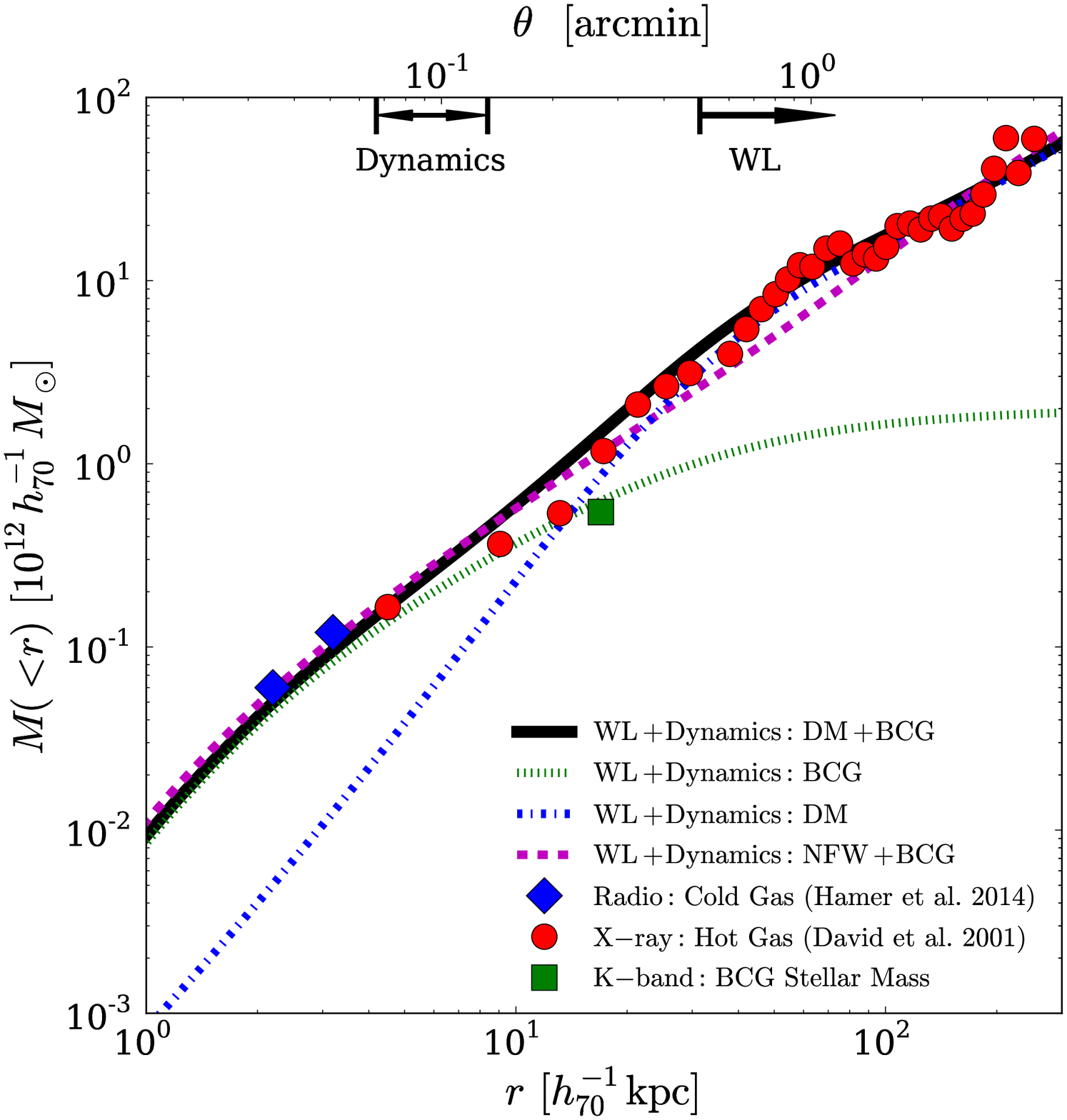}
\caption{Left: results for Hydra A including an additional extended dark-matter
 component associated with the BCG. 
The red-solid lines show the best-fit models for the total lensing
 signal and the projected velocity dispersion.
The green-dotted, magenta-dashed, and blue-dot-dashed lines are the
 lensing signals for the stellar component, the dark-matter component
 associated with the BCG, and the NFW halo component, respectively.
Right: spherical mass profile within $300$\,kpc.
The back-solid, green-dotted, and blue-dotted-dashed lines show the
 results for the total mass, stellar mass, and total dark-matter
 distributions, respectively. 
For comparison, 
the total mass profile in Figure \ref{fig:WL+dynamics} is shown by the magenta-dashed line.
The blue diamonds, red circles, and green squares are the same as those in
 Figure \ref{fig:WL+dynamics2}.}
\label{fig:WL+dynamics2}
\end{figure*}

\begin{figure}
\includegraphics[width=\hsize]{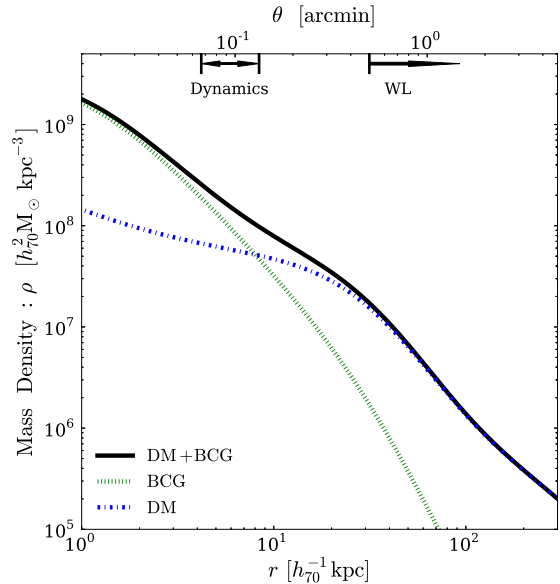}
\caption{Top: mass density profiles in the central region of Hydra A.
The black-solid, green-dashed, blue-dotted and red-dotted lines show 
the total, BCG stellar, dark-matter, and NFW mass density profiles,
 respectively. Bottom: mass density slope, $-d\ln \rho/d\ln r$.
The best-fit dark-matter density profile is shallower
 than the stellar density profile at $1 {\rm kpc}\simlt r\simlt20$kpc, 
as imposed by the prior and favored by the data.}
\label{fig:density}
\end{figure}

Figure \ref{fig:WL+dynamics2} shows the resulting best-fit tangential
shear 
profile $g_+(\theta)$ (left) and the spherical mass profile $M(<r)$ (right).
We find that adding the extended central dark-matter component can fully
account for the the dynamical and lensing data.
The change on the total mass profile due to the additional dark-matter
component is small (right panel of Figure \ref{fig:WL+dynamics2}). 
The resulting total mass profile agrees better with
those from independent measurements.
In Figure \ref{fig:density} we show the spherical mass density profiles
for the total, dark-matter, NFW and BCG
stellar components in the central region.
The dark-matter density profile at $1~{\rm kpc}\simlt r\simlt20$kpc 
is shallower than the stellar density profile, as imposed by the prior
(the bottom panel in Figure \ref{fig:density}).
The slope for the total matter density is perturbed by the BCG
stellar and the additional dark-matter components from the slope of the
NFW main halo.
The minimized $\chi^2_{\rm min} (\mathrm{d.o.f.})=6.1$ $(20)$ has been
improved from $9.0$ $(23)$ relative to the NFW model. 
The $F$ test gives a probability of $5\times10^{-2}$, thus supporting
the addition of the flat dark-matter component to explain better both lensing and kinematics data.

Lensing systematics, such as the effects of halo/BCG triaxiality
\citep[e.g.][]{Corless07,Oguri09,Meneghetti10,Umetsu15},
cannot also be ignored as one of the possible causes 
of the discrepancy between weak-lensing data and the best-fit
model for the NFW and BCG stellar mass components (Section \ref{subsec:joint_hydraA}).
Since the additional dark-matter component would not be a unique solution to
explain the discrepancy, it is also crucial to investigate other possibilities.
Sophisticated model of the triaxial/rotational/non-axisymmetric halo model 
for stellar kinematics and weak-lensing data would be critically important 
for a deeper understanding of the connection between the BCGs and host
dark matter halo.
Further systematic studies of very nearby clusters are of vital
importance for understanding of the mass distribution around BCGs.

\subsection{Dark-Matter Fraction around the BCG in Hydra A}\label{subsec:fDM}

Isobaric cooling time scales for the two clusters are less than $1$\,Gyr 
within the central $30$\,kpc \citep{David01,Sun03}.
Gas cooling thus causes condensation of baryons, leading to formation of
stars and
contraction of dark matter in the central cluster region.
One of the most plausible scenarios for the dark-matter response to the
cooling of baryons is adiabatic contraction
\citep[e.g.][]{Gnedin04}.
A joint analysis of stellar kinematics ($\simlt 8$kpc) and weak
lensing data  ($\simgt 30$kpc) for the Hydra A cluster allows us to 
decompose the observed total projected mass profile into the stellar and
dark-matter components around the BCG, which provides us with a good
opportunity to test the validity of adiabatic contraction models in
the central region.

Here we estimate dark-matter mass fractions around the BCG, 
given by $f_{\rm DM}=M_{\rm DM}/(M_{\rm DM}+M_{*}+M_{\rm gas})$, 
where $M_{\rm DM}$ is the dark-matter mass , $M_*$ is the
stellar mass, and $M_{\rm gas}$ is the gas mass
\citep{David01}.
We employ the NFW (Sections \ref{subsec:joint_a478} and
\ref{subsec:joint_hydraA}) and composite
dark-matter (NFW+PIEMD; Section \ref{subsec:joint_hydraA2})
models to calculate $M_{\rm DM}(<r)$.
The stellar mass is composed of the BCG and  cluster elliptical galaxies.
The BCG stellar mass profile is calculated using the best-fit
parameters given in Section \ref{subsec:joint_hydraA}. 
The stellar masses for other elliptical member galaxies 
are estimated by their $K$-band luminosities.
Here, we assume the Salpeter IMF (Section \ref{subsec:joint_hydraA})
because the stellar mass estimated by the Salpeter IMF
agrees with that estimated by the stellar kinematics.

\begin{figure}
\includegraphics[width=\hsize]{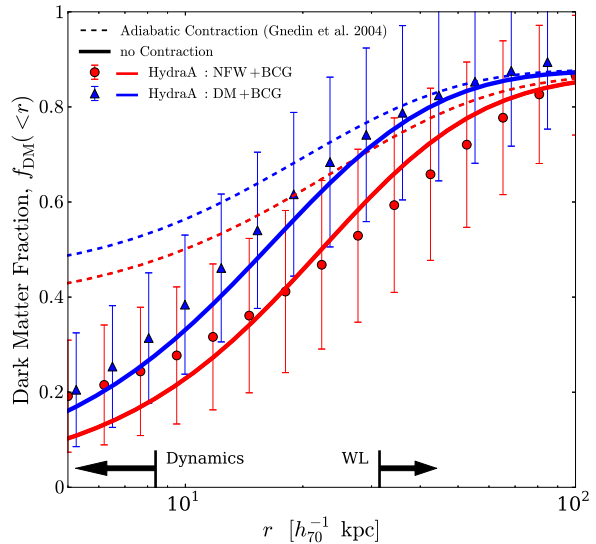}
\caption{Dark-matter fractions as a function of spherical cluster radius.
The red circles and blue triangles show the results for Hydra A using
 the best-fit NFW (Section \ref{subsec:joint_hydraA}) and 
NFW+PIEMD  (Section \ref{subsec:joint_hydraA2}) profiles, respectively. 
The solid and dashed lines are the dark-matter fractions predicted by
 models without 
and with adiabatic contraction, respectively \citep{Gnedin04}. The two
 arrows mark the radial range which covers each data.}
\label{fig:fDM}
\end{figure}

Figure \ref{fig:fDM} shows the resulting dark-matter fraction profiles 
for the Hydra A cluster.
The cumulative dark-matter fractions progressively increase with
increasing the cluster radius.
The slope is slightly steeper beyond the effective radii of BCGs,
$r_e\sim20{\rm kpc}$, and flattened at $r\sim 100$\,kpc.
Similar results have been reported by earlier work
\citep[e.g.][]{Nagino09,Oguri14}.
\cite{Nagino09} found that the integrated mass-to-light ratio profiles
for early-type galaxies based on the X-ray and photometry 
analysis increases beyond effective radii $r_e$ of galaxies.
\cite{Oguri14} conducted a joint analysis of strong lensing and stellar photometry 
for a large ensemble of elliptical galaxies, 
finding that the slope of the dark-matter faction 
becomes stepper at $r> r_e$.

Now we examine the dark-matter fractions using models with and without  
adiabatic contraction \citep{Gnedin04}.
We use the Jaffe model \citep{Jaffe83} as a stellar mass distribution,
where the density profile follows $\rho(r) \propto r^{-2}(r+s)^{-2}$.
In the limit of $r\gg a$ for the PIEMD model (Equation
(\ref{eq:PIEMD})), the PIEMD density profile resembles the Jaffe profile
For the dark-matter component, we only consider the NFW model.
The dark-matter fractions (dashed lines) predicted by adiabatic
contraction models \citep{Gnedin04} overestimate our measurements.
On the other hand, models without adiabatic contraction 
agree better with the dark-matter fractions, consistent with the results of 
\citet[][]{Newman13} and \citet{Oguri14}.
This suggests that other physical processes are critically important to 
suppress the modification of dark-matter profile.

\section{Discussion} \label{sec:dis2}

We find that the adiabatic contraction model \citep{Gnedin04}
overestimates the dark-matter fraction profiles compared to our measurements (Section \ref{subsec:fDM}).  
One of possible scenarios is that the gas outflows driven by AGN activities
suppress condensation of the dark-matter distribution in the central
region  \citep[e.g.][]{Teyssier11,Martizzi12,Ragone-Figueroa12}.
In fact, on-going jet activities of central AGNs have been reported in
two clusters \citep[e.g.]{David01,Sun03,Diehl08}. 
The projected distances between the BCG and X-ray bubbles triggered by
the observed activities are $r\sim25~{\rm kpc}$ for Hydra A and $r\sim9~{\rm kpc}$
for A478, respectively \citep[e.g.][]{David01,Sun03,Wise07,Diehl08}.
Three-dimentional characteristic scale radii 
obtained by equating the interior stellar and NFW masses
are $r_\mathrm{eq}\sim25~{\rm kpc}$ for Hydra A;
$r_\mathrm{eq}\sim22~{\rm kpc}$ and
$\sim15~{\rm kpc}$ for A478,
assuming the Salpeter and Chabrier IMFs, respectively.
This suggests that the current AGN jet activities can directly affect
the baryons over the region where stellar mass dominates.
Although there is an uncertainty between the three-dimentional
radius, $r_{\rm eq}$, and the projected distances between the BCG and X-ray bubbles, 
the results hold even if the jet/bubble direction is $\simlt 40^\circ$ 
from the plane of the sky.
Another alternative scenario assumes minor, dry (dissipationless) mergers
of clumpy structures \citep[e.g.][]{El-Zant01,Lackner10}, 
which can
flatten the central dark-matter density profile (Section
\ref{subsec:joint_hydraA2}) and be closely correlated with the BCG growth.

Further systematic studies of central 
dark-matter mass fractions and density slopes,
using a large sample of AGN and
non-AGN clusters, are required to 
distinguish between the effects of gas
outflows and dissipationless mergers,
for deep understanding of physical processes governing the BCG growth.
Systematic joint studies combining weak lensing and stellar kinematics
data will also be useful to examine correlations between stellar
population properties and IMF constraints inferred from joint analyses
\citep{McDermid14}.

\section{Summary} \label{sec:sum}

We have presented a weak-lensing study of the nearby cool-core clusters
Hydra A and A478. 
To minimize dilution of the lensing signal due to
contamination by cluster members, we carefully select populations of
background source galaxies in the color-magnitude plane.
Compared to weak-lensing studies of clusters at intermediate redshifts
$z\sim0.2$ \citep{Okabe13}, 
the level of background contamination by cluster members
is significantly reduced.
This enables us to obtain a larger number of
background galaxies for weak-lensing measurements while achieving a low
level of contamination. 
This is one of the advantages of weak-lensing analysis of very nearby clusters.

The S/N ratios for detection of the tangential shear signals  are $7.0$ and
$8.1$ for Hydra A and A478, respectively, after accounting for the
contribution from projected uncorrelated LSS.
The resulting S/N ratios are comparable to those of clusters at
$z\sim0.2$, thanks to the increased number of background galaxies behind
the clusters.

We find that the tangential shear profile for A478
is well fitted with a single NFW model.
For Hydra A, the tangential shear signal in the central region is
well described by a two-component model including the central BCG as an 
unresolved point mass.

We find that the choice of IMFs 
 can introduce a large uncertainty (factor of $\sim 2$) in the BCG
 stellar mass estimates, which makes it difficult to decompose the
 observed total projected mass profile into the stellar and dark-matter
 components (Section \ref{subsec:joint_a478}). 
On the other hand, as demonstrated by
\citet{Newman09,Newman11,Newman13}, the internal stellar kinematics of
BCGs enables us to
precisely measure the mass profiles of two components 
independent of the IMF uncertainty (Section \ref{subsec:joint_hydraA}).
For Hydra A, we find that
the central mass profile ($ < 300$\,kpc) determined from weak lensing is in excellent
agreement with those from independent measurements, including dynamical masses
estimated from the cold gas disk component, X-ray hydrostatic total mass estimates,
and central stellar mass estimates.
For the BCG, the data prefer the Salpeter IMF to the Chabrier and Kroupa
IMFs.
An additional flat dark-matter component around the BCG accounts
simultaneously for the weak-lensing and stellar-kinematics data
for Hydra A.

Dark-matter fractions around the BCGs for
the Hydra A cluster
are found to be smaller than those predicted by adiabatic contraction models, and
to agree well with model predictions without contraction.
This implies that other baryonic processes, such as the AGN feedback, 
dissipationless mergers, or the combination of the two, 
could play important roles in shaping the central cluster mass profile.
A precise joint measurement of the central cluster mass profile provides
us with complementary information about the dynamics of hot intracluster
gas, which will be directly probed by Astro-H.
Stellar kinematics data for the BCG of A478 is also important for
future works.

\section*{Acknowledgments}

We are grateful to N. Kaiser for developing the IMCAT package and making it publicly available. 
We thank Masamune Oguri, Masanori Nakamura, Takashi Hamana, Claire
Nicole Lackner and Alexie Leauthaud for helpful discussions.
This work was supported by ``World Premier International Research Center
Initiative (WPI Initiative)`` and the Funds for the Development of Human
Resources in Science and Technology under MEXT, Japan, and 
 Core Research for Energetic Universe in Hiroshima University (the MEXT
 program for promoting the enhancement of research universities, Japan).
N.Okabe (26800097), M. Takizawa (26400218), and K. Sato
(25800112) are supported by a Grant-in-Aid from the Ministry of Education, Culture, 
Sports, Science, and Technology of Japan.
K. Umetsu acknowledges partial support from 
the Ministry of Science and Technology of Taiwan
(grant MOST 103-2112-M-001-030-MY3).

\bibliographystyle{mn2e}
\bibliography{my,hydraa}

\appendix

\section{Multiwavelength Data of the Hydra A cluster}

We here summarize the observing data and their measurement errors 
for the tangential shear profile, the projected velocity dispersion and the
light profile for Hydra A.

\begin{table}
\caption{Tangential shear ($g_+$), projected velocity dispersion
 ($\sigma_{v,*}$) and the core (a) and scale  radii (s) of the PIEMD for the Hydra A. Note
 that the measurement error for the tangential shear ($C=C_{\rm stat}+C_{\rm LSS}$) is weakly correlated. $^\dagger$:\citet{Heckman85}.$^\ddagger$:\citet{Loubser08}.} \label{tab:hydraAdata}
\begin{center}
\begin{tabular}{ccc}
\hline
\hline
$\theta$ & $g_+$ & $C^{1/2}$ \\
(arcmin) & &  \\
\hline
$5.65922\times10^{-1}$ & $1.75256\times10^{-1}$ & $1.12041\times10^{-1}$ \\
$7.13965\times10^{-1}$ & $1.28711\times10^{-1}$ & $7.97313\times10^{-2}$ \\
$9.03514\times10^{-1}$ & $7.36090\times10^{-2}$ & $6.06401\times10^{-2}$ \\
$1.12847$ & $7.98743\times10^{-2}$ & $4.91149\times10^{-2}$ \\
$1.43289$ & $5.48124\times10^{-2}$ & $4.17630\times10^{-2}$ \\
$1.79821$ & $5.15705\times10^{-2}$ & $2.83264\times10^{-2}$ \\
$2.27188$ & $4.08415\times10^{-2}$ & $2.18615\times10^{-2}$ \\
$2.83350$ & $2.58124\times10^{-2}$ & $1.76992\times10^{-2}$ \\
$3.57655$ & $2.13949\times10^{-2}$ & $1.37130\times10^{-2}$ \\
$4.51228$ & $7.67107\times10^{-3}$ & $1.09288\times10^{-2}$ \\
$5.69914$ & $1.69769\times10^{-2}$ & $8.85021\times10^{-3}$ \\
$7.14476$ & $7.01793\times10^{-3}$ & $7.00017\times10^{-3}$ \\
$8.97936$ & $1.40250\times10^{-2}$ & $5.61649\times10^{-3}$ \\
$1.12947\times10$ & $1.04568\times10^{-2}$ & $4.69525\times10^{-3}$ \\
$1.42627\times10$ & $9.69863\times10^{-3}$ & $3.89583\times10^{-3}$ \\
$1.78611\times10$ & $7.65808\times10^{-3}$ & $3.38880\times10^{-3}$ \\
$2.25777\times10$ & $7.34974\times10^{-3}$ & $2.91867\times10^{-3}$ \\
$2.84133\times10$ & $4.07282\times10^{-3}$ & $2.49727\times10^{-3}$ \\
$3.56222\times10$ & $1.74820\times10^{-3}$ & $2.24314\times10^{-3}$ \\
$4.41888\times10$ & $2.57557\times10^{-3}$ & $2.20707\times10^{-3}$ \\

\hline
$\theta$ & $\sigma_{v,*}$ & $\delta_{\sigma_{v,*}}$ \\
(arcmin) & (kms$^{-1}$) & (kms$^{-1}$) \\
\hline
$6.83333\times10^{-2\dagger}$ & $3.29000\times10^{2}$ & $5.70000\times10$ \\
$7.00000\times10^{-2\dagger}$ & $3.83000\times10^{2}$ & $6.10000\times10$ \\
$8.03333\times10^{-2\ddagger}$ & $3.29000\times10^{2}$ & $2.60000\times10$ \\
$9.50000\times10^{-2\dagger}$ & $2.54000\times10^{2}$ & $1.18000\times10^{2}$ \\
$1.01667\times10^{-1\dagger}$ & $2.85000\times10^{2}$ & $8.20000\times10$ \\
$1.30000\times10^{-1\dagger}$ & $3.39000\times10^{2}$ & $1.13000\times10^{2}$ \\
\hline
$a$ & $s$&  \\
(arcmin) & (arcmin) &  \\
\hline
$1.751265\times10^{-2}$ & $4.68178\times10^{-1}$ \\
\hline
\end{tabular}
\end{center} 
\end{table}

\bsp

\label{lastpage}

\end{document}